\documentclass[amsmath,amsfonts,amssymb]{revtex4} 
\usepackage{stmaryrd}
\usepackage{array}
\usepackage{mathrsfs}

\newtheorem{theorem}{Theorem}
\newtheorem{corollary}{Corollary}
\newtheorem{lemma}{Lemma}
\newtheorem{remark}{Remark}

\newenvironment{proof}{\noindent\scshape Proof\normalfont.}{\hfill$\Box$\medskip}

\newcommand{\const}{\mathrm{const}}

\newcommand{\veps}{\varepsilon}
\newcommand{\osym}{o}
\newcommand{\Osym}{\mathcal{O}}
\renewcommand{\d}{\mathrm{d}}

\newcommand{\Z}{\mathbb{Z}}
\newcommand{\Q}{\mathbb{Q}}
\newcommand{\R}{\mathbb{R}}
\newcommand{\C}{\mathbb{C}}
\newcommand{\MzC}{\mathrm{M_{2}(\C)}}
\newcommand{\Hyp}[1]{{\upshape (#1)}}

\newcommand{\sign}{\mathrm{sgn}}
\newcommand{\diag}{\mathrm{diag}\,}
\newcommand{\rank}{\mathrm{rank}\,}

\newcommand{\re}{\mathrm{Re}\,}

\newcommand{\sdot}{\,\cdot\,}
\newcommand{\half}{\frac{1}{2}}

\newcommand{\hs}{\mathscr{L}}

\newcounter{num}
\newenvironment{clist}[1]{\begin{list}{}%
{\renewcommand{\makelabel}%
 {\stepcounter{num}\hfill\mbox{\upshape\thenum}}%
 \setcounter{num}{0}%
 \settowidth{\labelwidth}{\upshape #1}%
 \setlength{\labelsep}{0.5em}%
 \setlength{\leftmargin}{\labelwidth}%
 \addtolength{\leftmargin}{\labelsep}%
 \setlength{\itemsep}{0.5ex}%
}}{\end{list}}

\newenvironment{plist}
{\begin{list}{}{
\setlength{\leftmargin}{2.5em}
\setlength{\labelwidth}{2em}
\setlength{\itemsep}{0.5ex}
}}{\end{list}}

\begin{document}

\title{On the Eigenvalues of the Chandrasekhar-Page Angular Equation}

\author{Davide Batic}
 \email{davide.batic@mathematik.uni-regensburg.de}
\author{Harald Schmid}
 \email{harald.schmid@mathematik.uni-regensburg.de}
\affiliation{NWF I - Mathematik, Universit{\"a}t Regensburg, D-93040 Regensburg, Germany.}
\author{Monika Winklmeier}
 \email{winklmeier@math.uni-bremen.de}
\affiliation{FB 3-Mathematik, Universit{\"a}t Bremen, D-28359 Bremen, Germany.}


\begin{abstract}\noindent
In this paper we study for a given azimuthal quantum number $\kappa$ the eigenvalues of the Chandrasekhar-Page angular
equation with respect to the parameters $\mu:=am$ and $\nu:=a\omega$, where $a$ is the angular momentum per unit mass of
a black hole, $m$ is the rest mass of the Dirac particle and $\omega$ is the energy of the particle (as measured at
infinity). For this purpose, a self-adjoint holomorphic operator family $A(\kappa;\mu,\nu)$ associated to this eigenvalue
problem is considered. At first we prove that for fixed $\kappa\in\R\setminus(-\frac{1}{2},\frac{1}{2})$ the spectrum of
$A(\kappa;\mu,\nu)$ is discrete and that its eigenvalues depend analytically on $(\mu,\nu)\in\C^2$. Moreover, it will be
shown that the eigenvalues satisfy a first order partial differential equation with respect to $\mu$ and $\nu$, whose characteristic
equations can be reduced to a Painlev{\'e} III equation. In addition, we derive a power series expansion for
the eigenvalues in terms of $\nu-\mu$ and $\nu+\mu$, and we give a recurrence relation for their coefficients. Further,
it will be proved that for fixed $(\mu,\nu)\in\C^2$ the eigenvalues of $A(\kappa;\mu,\nu)$ are the zeros of a holomorphic
function $\Theta$ which is defined by a relatively simple limit formula. Finally, we discuss the problem if there exists
a closed expression for the eigenvalues of the Chandrasekhar-Page angular equation.
\end{abstract}

\maketitle

\section{Introduction}

The angular eigenvalue problem of a spin-$\half$ particle in the Kerr-Newman geometry is given by the
Chandrasekhar-Page angular equation (see \cite[Chap. 10, Sec. 104]{Chandra2})
\begin{align}
\mathcal{L}^{+}_{\half}S_{+\half} & = \left(am\cos\theta-\lambda\right)S_{-\half}, \label{CPE1} \\
\mathcal{L}^{-}_{\half}S_{-\half} & = \left(am\cos\theta+\lambda\right)S_{+\half}, \label{CPE2}
\end{align}
where the Kerr parameter $a$ is the angular momentum per unit mass of a black hole and $m$ is the rest mass of the Dirac
particle. Moreover, the differential operators $\mathcal{L}^{\pm}_{\half}$ are defined by
\begin{equation*}
\mathcal{L}^{\pm}_{\half} = \partial_{\theta}\pm Q(\theta)+\frac{\cot\theta}{2}\,,\quad
Q(\theta) := a\omega\sin\theta+\frac{\kappa}{\sin\theta}\,,\quad \theta\in(0,\pi),
\end{equation*}
where $\omega$ is the energy of the particle (as measured at infinity) and $\kappa$ is a half-integer, i.e.,
$\kappa=k-\half$ with some $k\in\Z$. A parameter $\lambda\in\R$ is called an \emph{eigenvalue} of this spectral problem
if the system given by \eqref{CPE1} -- \eqref{CPE2} has a nontrivial solution which is square-integrable on $(0,\pi)$ with
respect to the weight function $\sin\theta$. In this paper we study for fixed $\kappa$ the eigenvalues of the
Chandrasekhar-Page angular equation as a function of the parameters $\mu:=am$ and $\nu := a\omega$. As a main result, we
will prove that the eigenvalues satisfy a first order quasi-linear partial differential equation, and we will derive a
power series expansion for the eigenvalues in terms of $\nu-\mu$ and $\nu+\mu$.

For this purpose it is necessary to consider the system \eqref{CPE1} -- \eqref{CPE2} in a more general context where
$\kappa$ is real, $|\kappa|\geq\frac{1}{2}$, and $\mu$, $\nu$ are complex numbers. At first we rewrite this system
for fixed $\kappa\in\R\setminus(-\frac{1}{2},\frac{1}{2})$ as an eigenvalue problem for some self-adjoint holomorphic
operator family $A=A(\kappa;\mu,\nu)$ depending on the parameters $(\mu,\nu)\in\C^2$. In the special case where
$(\mu,\nu)\in\R^2$ the differential operator $A(\kappa;\mu,\nu)$ is self-adjoint and has purely discrete spectrum. In Section
II we prove that for a given $\kappa$ the eigenvalues $\lambda_j(\kappa;\mu,\nu)$ of $A$ are holomorphic functions in
$(\mu,\nu)$, and we derive some basic estimates for them. Furthermore, we transform the system
\eqref{CPE1} -- \eqref{CPE2} to a matrix differential equation
\begin{equation} \label{AngSys0}
y'(x) = \left[\frac{1}{x}\,B_0 + \frac{1}{x-1}\,B_1 + C\right]\,y(x)
\end{equation}
on the interval $(0,1)$ with coefficient matrices
\begin{equation*}
B_0 = \left(\begin{array}{cc} -\frac{\kappa}{2}-\frac{1}{4} & \mu-\lambda \\[1ex] 0 & \frac{\kappa}{2}+\frac{1}{4} \end{array}\right),\quad
B_1 = \left(\begin{array}{cc} \frac{\kappa}{2}+\frac{1}{4} & 0 \\[1ex] \mu-\lambda & -\frac{\kappa}{2}-\frac{1}{4} \end{array}\right),\quad
 C  = \left(\begin{array}{cc} -2\nu & -2\mu \\[1ex] 2\mu & 2\nu \end{array}\right),
\end{equation*}
which can be extended to the complex domain $\C\setminus\{0,1\}$. In this way we obtain a further characterisation
of the eigenvalues of $A$ and some useful estimates for the corresponding eigenfunctions. Applying analytic perturbation
theory, we show in Section III that the eigenvalues $\lambda_j(\kappa;\mu,\nu)$ satisfy the partial differential equation
\begin{equation} \label{PDE0}
\left(\mu-2\,\nu\,\lambda\right)\frac{\partial\lambda}{\partial\mu} +
\left(\nu-2\,\mu\,\lambda\right)\frac{\partial\lambda}{\partial\nu} + 2\,\kappa\,\mu + 2\,\mu\,\nu = 0.
\end{equation}
In particular, this result can be used to obtain a recurrence relation for the coefficients $c_{m,n}$ of a power series
expansion
\begin{equation*}
\lambda_j(\kappa;\mu,\nu) = \sum_{m,n=0}^{\infty}c_{m,n}(\nu-\mu)^m(\nu+\mu)^n.
\end{equation*}
In Section IV we solve the PDE \eqref{PDE0} by the method of characteristics. First, we derive an explicit formula for
the eigenvalues in the case $|\mu|=|\nu|$. Moreover, in the regions where $|\mu|\neq|\nu|$ we reduce
the characteristic equations of \eqref{PDE0} to a Painlev{\'e} III equation
\begin{equation*}
v\,v' + t\,v\,v'' - t(v')^2 - 2\,\kappa\left(v^2\pm 1\right)v - t\left(v^4-1\right) = 0
\end{equation*}
with parameters $\alpha=\pm\beta=2\kappa$ and $\gamma=-\delta=1$ according to the notation in \cite{MCB} and \cite{MW}.
As this differential equation is in general not solvable in terms of elementary functions, we cannot expect a closed
expression for the eigenvalues of the Chandrasekhar-Page angular equation for all $(\mu,\nu)\in\R^2$. However, if
$\kappa$ is a half-integer, i.e., $\kappa=k-\frac{1}{2}$ with some positive integer $k$, then $\alpha\pm\beta=2(2k-1)$,
and there are integrals of polynomial type for the third Painlev{\'e}
equation in this special case, cf. \cite{MCB}. Hence, if $\kappa=\pm\frac{1}{2},\pm\frac{3}{2},\ldots$, there exist algebraic solutions
of the partial differential equation \eqref{PDE0}, and the question arises if these explicit solutions are in fact
eigenvalues of the Chandrasekhar-Page angular equation. It turns out that there is another type of ``special values"
associated to the operator $A$, called \emph{monodromy eigenvalues}, which belong to the algebraic solutions of the PDE
\eqref{PDE0}. For a half-integer $\kappa$, the monodromy eigenvalues are introduced in Section V by requiring that the
system
\eqref{AngSys0} has a fundamental matrix of the form
\begin{equation*}
[x(1-x)]^{-\frac{\kappa}{2}-\frac{1}{4}}H(x)
\end{equation*}
with an entire matrix function $H:\C\longrightarrow\MzC$. This property turns out to be equivalent to the existence of
special solutions of the form
\begin{equation*}
[x(1-x)]^{-\frac{\kappa}{2}-\frac{1}{4}}p^{\pm}(x)e^{\pm 2tx},
\end{equation*}
where $p^{\pm}:\C\longrightarrow\C^2$ are polynomials and $t = \pm\sqrt{\nu^2-\mu^2}$. For comparison purposes, an
eigenvalue of $A$ can be characterised by the property that \eqref{AngSys0} possesses a nontrivial solution of the form
\begin{equation*}
[x(1-x)]^{\frac{\kappa}{2}+\frac{1}{4}}\eta(x)
\end{equation*}
with some entire vector function $\eta:\C\longrightarrow\C^2$. We prove that the monodromy eigenvalues are zeros of a
polynomial with degree $2k-1$ whose coefficients are polynomials in $\mu$ and $\nu$. Moreover, it can be shown that
monodromy eigenvalues and ``classical" eigenvalues are distinct at least in a neighbourhood of $(\mu,\nu)=(0,0)$.
Nevertheless, they are both characterised by the fact that certain monodromy data of the system \eqref{AngSys0} are
preserved for all parameters $(\mu,\nu)$. In fact, $\lambda$ is a monodromy eigenvalue of $A$ if and only if the
monodromy matrices of \eqref{AngSys0} at the regular-singular points $0$ and $1$ are diagonal, whereas $\lambda$ is a
classical eigenvalue of $A$ if and only if a certain non-diagonal entry of the connection matrix for the fundamental
matrices at $0$ and $1$ vanishes. Hence, for the Chandrasekhar-Page angular equation the monodromy as well as the
classical eigenvalue problem is closely related to the isomonodromy problem for the differential equation \eqref{AngSys0}.
Monodromy
preserving deformations for such a system were studied by Jimbo, Miwa \& Ueno in \cite{JMU1}, however, only for the case
that the eigenvalues of $B_0$ and $B_1$ do not differ by an integer, i.e., $\kappa+\frac{1}{2}\not\in\Z$. In Section VI
we consider the isomonodromy problem for \eqref{AngSys0} in the case that $\kappa$ is a half-integer. As a consequence,
we show that the monodromy eigenvalues of $A$ satisfy the partial differential equation \eqref{PDE0}, and we obtain an
alternative derivation of \eqref{PDE0} for the classical eigenvalues of $A$. Unlike the proof in Section III, which
relies on the particular structure of the Chandrasekhar-Page angular equation, the method presented in Section V is more
general and based on finding suitable deformation equations for parameter-dependent differential equations. Thus, we
expect that this technique is applicable to other eigenvalue problems as well.

\section{A self-adjoint holomorphic operator family associated to the Chandrasekhar-Page angular equation}

By introducing the notations
\begin{equation*}
\mu := am,\quad \nu := a\omega,\quad
S(\theta) := \sqrt{\sin\theta}\left(\begin{array}{c} S_{+\half}(\theta) \\[1ex] S_{-\half}(\theta) \end{array}\right),
\end{equation*}
the Chandrasekhar-Page angular equation \eqref{CPE1} -- \eqref{CPE2} takes the form
\begin{equation} \label{AngEqu}
(\mathfrak{A}\,S)(\theta) := \left(\begin{array}{cc} 0 & 1 \\[1ex] -1 & 0 \end{array}\right)S'(\theta)
+ \left(\begin{array}{cc}
-\mu\cos\theta & -\frac{\kappa}{\sin\theta}-\nu\sin\theta \\[1ex]
-\frac{\kappa}{\sin\theta}-\nu\sin\theta & \mu\cos\theta \end{array}\right)S(\theta) = \lambda\,S(\theta),\quad\theta\in(0,\pi),
\end{equation}
with fixed $\kappa\in\R\setminus(-\frac{1}{2},\frac{1}{2})$ and parameters $(\mu,\nu)\in\C^2$.
We can associate the so called  minimal operator $A_0$ to the formal differential expression $\mathfrak A$, which acts
in the Hilbert space $\mathcal{H} := \hs^2\left((0,\pi), \C^{2}\right)$ of square integrable vector functions with
respect to the scalar product
\begin{equation} \label{Prod}
(S_1,S_2) := \int_{0}^{\pi} S_2(\theta)^{\ast}S_1(\theta)\,\d\theta,\quad S_1,\,S_2\in\mathcal{H}.
\end{equation}
The operator $A_0$  given by $\mathcal{D}(A_0)=\mathrm{C_0^\infty}\left((0,\pi),\C^{2}\right)$ and
$A_0 S := \mathfrak{A}\,S$ for $S\in\mathcal{D}(A_0)$ is densely defined and closable.
For $|\kappa|\ge \frac{1}{2}$ and $(\mu,\nu)\in\R^2$ the formal differential operator in \eqref{AngEqu} is in the limit
point case at $0$ and $\pi$, hence $A_0$ is even essentially self-adjoint. In the following we
denote the closure of $A_0$ by $A=A(\kappa;\mu,\nu)$. According to \cite[Theorem 5.8]{Weidmann} the domain of
$A(\kappa;0,0)$ is given by
\begin{equation*}
\mathcal{D}(A)=\left\{S\in\mathcal{H}\,:\,S\mbox{ is absolutely continuous and }A(\kappa;0,0)S\in\mathcal{H}\right\}.
\end{equation*}
Since $A(\kappa;\mu,\nu)=A(\kappa;0,0)+T(\mu,\nu)$ with the bounded multiplication operator
\begin{equation*}
T(\mu,\nu) = \left(\begin{array}{cc}
-\mu\cos\theta & -\nu\sin\theta \\[1ex] -\nu\sin\theta & \mu\cos\theta \end{array}\right),
\end{equation*}
its domain of definition $\mathcal{D}(A)$ is independent of $(\mu,\nu)\in\C^2$ (see
\cite[Chap. IV, \S\,1, Theorem 1.1]{Kato}).
Moreover, if $(\mu,\nu)\in\R^2$, then $T(\mu,\nu)$ is a symmetric perturbation of $A(\kappa;0,0)$, and
\cite[Chap. V, \S\,4, Theorem 4.10]{Kato} yields that $A(\kappa;\mu,\nu)$ is self-adjoint. Thus,
according to Kato's classification \cite[Chap. VII, \S\,3]{Kato}, $A(\kappa;\mu,\nu)$ forms a self-adjoint holomorphic
operator family of type (A) in the variables $(\mu,\nu)\in\C^2$. Further, the spectrum of $A(\kappa;0,0)$ is discrete and
consists of simple eigenvalues given by
\begin{equation} \label{Init}
\lambda_j(\kappa;0,0) = \sign(j)\left(|\kappa|-\half+|j|\right),\quad j\in\Z\setminus\{0\}
\end{equation}
(for the details we refer to Appendix A). This means, in particular, that $A(\kappa;0,0)$ has compact resolvent, and
from \cite[Chap. V, \S\,2, Theorem 2.4]{Kato} it follows that $A(\kappa;\mu,\nu)$ has compact resolvent for all
$(\mu,\nu)\in\C^2$. As a consequence, the spectrum of $A(\kappa;\mu,\nu)$, $(\mu,\nu)\in\C^2$, is discrete, and since
$A(\kappa;\mu,\nu)$ is in the limit point case at $\theta=0$ and $\theta=\pi$, it consists of simple eigenvalues for
$(\mu,\nu)\in\R^2$. Now, \cite[Chap. V, \S\,3, Theorem 3.9]{Kato} implies that the eigenvalues
$\lambda_j=\lambda_j(\kappa;\mu,\nu)$, $j\in\Z\setminus\{0\}$, of $A(\kappa;\mu,\nu)$ are simple and depend
holomorphically on $(\mu,\nu)$ in a complex neighbourhood of $\R^2$. Moreover, the partial derivatives of $A$ with respect
to $\mu$ and $\nu$ are given by
\begin{equation*}
\frac{\partial A}{\partial\mu} = \left(\begin{array}{cc}
-\cos\theta & 0 \\[1ex] 0 & \cos\theta \end{array}\right),\quad
\frac{\partial A}{\partial\nu} = \left(\begin{array}{cc}
0 & -\sin\theta \\[1ex] -\sin\theta & 0 \end{array}\right),
\end{equation*}
which yields the following estimates for the growth rate of the eigenvalues (compare \cite[Chap. VII, \S 3, Sec. 4]{Kato}):
\begin{equation*}
\left|\frac{\partial \lambda_j}{\partial\mu}\right|\leq\left\|\frac{\partial A}{\partial\mu}\right\|\leq 1,\quad
\left|\frac{\partial \lambda_j}{\partial\nu}\right|\leq\left\|\frac{\partial A}{\partial\nu}\right\|\leq 1.
\end{equation*}
Here, $\|\cdot\|$ denotes the operator norm of a $(2\times 2)$ matrix. In addition, by
\cite[Chap. V, \S\,3, Theorem 4.10]{Kato}, we have
\begin{equation} \label{Dist}
\min_{j\in\Z\setminus\{0\}}\left|\lambda-\lambda_j(\kappa;0,0)\right|\leq\|T(\mu,\nu)\|\leq\max\{|\mu|,|\nu|\}
\end{equation}
for each eigenvalue $\lambda$ of $A(\kappa;\mu,\nu)$. Finally, by interchanging the components of $S(\theta)$, we
obtain that a point $\lambda$ is an eigenvalue of $A(\kappa;\mu,\nu)$ if and only if $-\lambda$ is an eigenvalue of
$A(-\kappa;\mu,-\nu)$. Since the eigenvalues depend holomorphically on $\mu$ and $\nu$, the identity
\begin{equation*}
\lambda_j(\kappa;\mu,\nu) = -\lambda_{-j}(-\kappa;\mu,-\nu)
\end{equation*}
holds for all $(\mu,\nu)$ in a neighbourhood of $\R^2$. Therefore, we restrict our attention to the case
$\kappa\in[\frac{1}{2},\infty)$. Note that $\lambda\in\C$ is an eigenvalue of $A(\kappa;\mu,\nu)$ if and only if
the system \eqref{AngEqu} has a nontrivial solution $S(\theta)$ satisfying
\begin{equation} \label{AngInt}
\int_{0}^{\pi} |S(\theta)|^2\,\d\theta < \infty.
\end{equation}
By means of the transformation
\begin{equation} \label{Trafo1}
S(\theta) = \left(\begin{array}{cc} \sqrt{\tan\frac{\theta}{2}} & 0 \\[1ex]
0 & \sqrt{\cot\frac{\theta}{2}} \end{array}\right)y({\textstyle\sin^2\frac{\theta}{2}}),\quad\theta\in(0,\pi),
\end{equation}
the differential equation \eqref{AngEqu} is equivalent to the system
\begin{equation} \label{AngSys}
y'(x) = \left[\frac{1}{x}\,B_0 + \frac{1}{x-1}\,B_1 + C\right]\,y(x)
\end{equation}
on the interval $(0,1)$ with coefficient matrices
\begin{equation} \label{Coeff1}
B_0 := \left(\begin{array}{cc} -\frac{\kappa}{2}-\frac{1}{4} & \mu-\lambda \\[1ex] 0 & \frac{\kappa}{2}+\frac{1}{4} \end{array}\right),\quad
B_1 := \left(\begin{array}{cc} \frac{\kappa}{2}+\frac{1}{4} & 0 \\[1ex] \mu-\lambda & -\frac{\kappa}{2}-\frac{1}{4} \end{array}\right),\quad
 C  := \left(\begin{array}{cc} -2\nu & -2\mu \\[1ex] 2\mu & 2\nu \end{array}\right),
\end{equation}
and the normalisation condition \eqref{AngInt} becomes
\begin{equation} \label{SysInt}
\int_{0}^{1} y(x)^\ast\left(\begin{array}{cc} \frac{1}{1-x} & 0 \\[1ex] 0 & \frac{1}{x} \end{array}\right)y(x)\,\d x < \infty.
\end{equation}
If we consider the differential equation \eqref{AngSys} for a fixed $\kappa\in(0,\infty)$ in the complex plane, then it
has two regular singular points, one at $x=0$ and one at $x=1$ with characteristic values
$\pm\left(\frac{\kappa}{2}+\frac{1}{4}\right)$. From the theory of asymptotic expansions (see \cite{Wasow}, for example),
it follows that for each $\lambda\in\C$ there exists a nontrivial solution
\begin{equation} \label{Holom}
y_0(x,\lambda) = x^{\frac{\kappa}{2}+\frac{1}{4}}h(x,\lambda),\quad x\in\mathfrak{B}_0,
\end{equation}
of \eqref{AngSys} in the unit disc $\mathfrak{B}_0\subset\C$ with centre $0$, where
$h(\sdot,\lambda):\mathfrak{B}_0\longrightarrow\C^2$ is a holomorphic function,
\begin{equation} \label{Funct}
h(x,\lambda) = \sum_{n=0}^\infty x^n h_n(\lambda),\quad h_0(\lambda) := \left(\begin{array}{c} \mu-\lambda \\[1ex] \kappa+\frac{1}{2} \end{array}\right).
\end{equation}
Here $h_0(\lambda)$ is an eigenvector of $B_0$ for the eigenvalue $\frac{\kappa}{2}+\frac{1}{4}$, and the coefficients
$h_n(\lambda)$, $n>1$, are uniquely determined by the recurrence relation
\begin{equation} \label{Recur}
\left(B_0-\alpha-n\right)h_n(\lambda) = \left(B_0+B_1-C+1-\alpha-n\right)h_{n-1}(\lambda) + C\,h_{n-2}(\lambda)
\end{equation}
with $\alpha := \frac{\kappa}{2}+\frac{1}{4}$ and $h_{-1}(\lambda):=0$. Since the matrices $B_0$ and $B_1$ depend
holomorphically on $\lambda$, the coefficients $h_n:\C\longrightarrow\C^2$ are holomorphic functions. By slightly
modifying the proof of \cite[Theorem 5.3]{Wasow}, it can be shown that the series \eqref{Funct} converges uniformly
in every compact subset of $\mathfrak{B}_0\times\C$. Thus, by a theorem of Weierstrass,
$h:\mathfrak{B}_0\times\C\longrightarrow\C^2$ is a holomorphic vector function in the variables $(x,\lambda)$. Now, let
\begin{equation*}
h\left(\textstyle{\frac{1}{2}},\lambda\right) =: \left(\begin{array}{c} f(\lambda) \\[1ex] g(\lambda) \end{array}\right),
\end{equation*}
and we define the holomorphic function $\Delta:\C\longrightarrow\C$ by
\begin{equation} \label{Delta}
\Delta(\lambda) := f(\lambda)^2-g(\lambda)^2,\quad\lambda\in\C.
\end{equation}
The following Lemma provides a connection between the eigenvalues of $A$ and the zeros of $\Delta$.

\begin{lemma} \label{Zeros1}
For fixed $\kappa\in[\frac{1}{2},\infty)$ and $(\mu,\nu)\in\C^2$, a point $\lambda\in\C$ is an eigenvalue of
$A(\kappa;\mu,\nu)$ if and only if $\lambda$ is a zero of the function $\Delta$ given by \eqref{Delta}. This is
equivalent to the statement that the differential equation \eqref{AngSys} has a nontrivial solution of the form
\begin{equation} \label{EigSol}
y(x) = [x(1-x)]^{\frac{\kappa}{2}+\frac{1}{4}}\eta(x),\quad x\in\C\setminus\{0,1\},
\end{equation}
where $\eta:\C\longrightarrow\C^2$ is an entire vector function. As a consequence, if $S$ is an eigenfunction of
$A(\kappa;\mu,\nu)$ for some eigenvalue $\lambda$, then
\begin{equation} \label{EigEst}
|S(\theta)|\leq C\sin^{\kappa}\theta,\quad \theta\in(0,\pi),
\end{equation}
with some constant $C>0$.
\end{lemma}

\begin{proof}
Defining
\begin{equation} \label{K}
K := \left(\begin{array}{cc} 0 & 1 \\[1ex] 1 & 0 \end{array}\right),
\end{equation}
we have $K^{-1} = K$ and $K B_0 K=B_1$, $K C K=-C$. Hence, $y$ is a solution of the system \eqref{AngSys} if and only if
the function $Ky(1-x)$ satisfies \eqref{AngSys}. In particular, $y_1(x) := K y_0(1-x)$ is a solution of \eqref{AngSys}
in the unit disc $\mathfrak{B}_1\subset\C$ with centre $1$, and $y_1$ has the form
\begin{equation*}
y_1(x,\lambda) = (1-x)^{\frac{\kappa}{2}+\frac{1}{4}}Kh(1-x,\lambda),\quad x\in\mathfrak{B}_1.
\end{equation*}
Moreover, by the Levinson Theorem (see \cite[Theorem 1.3.1]{Eastham}), any solution of \eqref{AngSys} which is linearly
independent of $y_0$ in $(0,1)$ behaves asymptotically like $x^{-\frac{\kappa}{2}-\frac{1}{4}}[v_0+\osym(1)]$ as $x\to 0$,
where $v_0$ is an eigenvector of $B_0$ for the eigenvalue $-\frac{\kappa}{2}-\frac{1}{4}$. Similarly, any solution of
\eqref{AngSys} which is linearly independent of $y_1$ in $(0,1)$ has the asymptotic behaviour
$(x-1)^{-\frac{\kappa}{2}-\frac{1}{4}}[v_1+\osym(1)]$ as $x\to 1$ with an eigenvector $v_1$ of $B_1$ for the eigenvalue
$-\frac{\kappa}{2}-\frac{1}{4}$. Now, if $\lambda$ is an eigenvalue of $A(\kappa;\mu,\nu)$, then the system \eqref{AngSys}
has a nontrivial solution $y$ satisfying \eqref{SysInt}, and it follows that $y(x) = y_a(x,\lambda)c_a$ holds in $(0,1)$
with some constants $c_a\in\C\setminus\{0\}$, $a\in\{0,1\}$. Thus, $y_0$ and $y_1$ are linearly dependent, and the
Wronskian $W(x,\lambda) := \det\left(y_0(x,\lambda),y_1(x,\lambda)\right)$ vanishes identically for all $x\in(0,1)$. In
particular, $0 = W\left(\frac{1}{2},\lambda\right) = 2^{-\kappa-\frac{1}{2}}\Delta(\lambda)$. Conversely, if
$\Delta(\lambda)=0$, then $W\left(\frac{1}{2},\lambda\right)=0$, which implies that $y_0$ and $y_1$ are linearly
dependent. Hence, $y_0(x) = y_1(x)c$ with some constant $c\in\C\setminus\{0\}$, and therefore $y_0$ is a solution of
\eqref{AngSys} satisfying the condition \eqref{SysInt} on the interval $(0,1)$. Moreover, we immediately obtain that
$y_0$ has the form \eqref{EigSol} with a holomorphic vector function
$\eta:\mathfrak{B}_0\cup\mathfrak{B}_1\longrightarrow\C^2$, and since
\eqref{AngSys} is regular in $\C\setminus\{0,1\}$, we can extend $\eta:\C\longrightarrow\C^2$ to an entire function by
the existence and uniqueness theorem. Finally, by means of the transformation \eqref{Trafo1}, an eigenfunction $S$ of
$A(\kappa;\mu,\nu)$ has to be a constant multiple of
\begin{equation*}
\sin^{\kappa}\theta\left(\begin{array}{cc} \sin\frac{\theta}{2} & 0 \\[1ex] 0 & \cos\frac{\theta}{2} \end{array}\right)\eta({\textstyle\sin^2\frac{\theta}{2}}),\quad\theta\in(0,\pi),
\end{equation*}
and this yields the estimate \eqref{EigEst}.
\end{proof}

\begin{lemma}
For fixed $\kappa\in[\frac{1}{2},\infty)$ and $j\in\Z\setminus\{0\}$, the $j$-th eigenvalue $\lambda_j(\kappa;\mu,\nu)$
of $A(\kappa;\mu,\nu)$ has a power series expansion of the form
\begin{equation} \label{Power}
\lambda_j(\kappa;\mu,\nu) = \sum_{m,n=0}^\infty \lambda_{m,n}\,\mu^m\nu^n,\quad \lambda_{0,0}=\lambda_j(\kappa;0,0),
\end{equation}
which is uniformly convergent in the polydisc $\mathfrak{C}:=\{(\mu,\nu)\in\C^2:|\mu|,\,|\nu|\leq\frac{1}{2}\}$.
Moreover, for all integers $m$ and $n$, the following estimate holds:
\begin{equation} \label{Coeff2}
\left|\lambda_{m,n}\right| \leq (|\kappa|+|j|)2^{n+m}.
\end{equation}
\end{lemma}

\begin{proof}
Since the coefficient matrices in \eqref{AngSys} depend holomorphically on $(\lambda,\mu,\nu)\in\C^3$, we can modify
\cite[Theorem 5.3]{Wasow} appropriately in order to obtain that $h$ in \eqref{Holom} and therefore
$\Delta=\Delta(\lambda,\mu,\nu)$ as given by \eqref{Delta} are holomorphic functions on $\C^3$. By a similar
reasoning as in the proof of Lemma \ref{Zeros1}, we can show that for fixed $(\mu,\nu)\in\C^2$ the eigenvalues of
$A(\kappa;\mu,\nu)$ coincide with the zeros of the function $\lambda\longmapsto\Delta(\lambda,\mu,\nu)$. In particular
for the case $(\mu,\nu)\in\R^2$ these zeros are simple because $A(\kappa;\mu,\nu)$ has only simple eigenvalues. Hence,
by solving the equation $\Delta(\lambda,\mu,\nu)=0$ and using the implicit function theorem, an eigenvalue
$\lambda_j(\kappa;\mu,\nu)$ of the operator $A(\kappa;\mu,\nu)$ depends holomorphically on $(\mu,\nu)$ in a complex
neighbourhood of $\R^2$. Furthermore, the estimate \eqref{Dist} implies that the set
$\left\{\lambda\in\C:\min_{j\neq 0}|\lambda-\lambda_j(\kappa;0,0)|\geq\frac{1}{2}\right\}$ contains
no eigenvalues of $A(\kappa;\mu,\nu)$ for all $(\mu,\nu)\in\mathfrak{C}$. Thus there exists a holomorphic solution
$\lambda:\mathfrak{C}\longrightarrow\C$ of the equation $\Delta(\lambda,\mu,\nu)=0$, which is uniquely determined by
$\lambda(0,0)=\lambda_j(\kappa;0,0)$. Consequently, $\lambda_j(\kappa;\mu,\nu)$ is holomorphic
in $\mathfrak{C}$, and therefore it has a power series expansion in $\mathfrak{C}$ of the form \eqref{Power}. In
addition, by Cauchy's formula,
\begin{equation*}
\lambda_{m,n} = -\frac{1}{4\pi^2}\oint_{\partial\mathfrak{C}}\frac{\lambda_j(\kappa;\mu,\nu)}{\mu^{m+1}\nu^{n+1}}\,\d\mu\,\d\nu,
\end{equation*}
and applying \eqref{Dist} and \eqref{Init}, it follows that
\begin{equation*}
\left|\lambda_j(\kappa;\mu,\nu)\right|\leq\left|\lambda_j(\kappa;0,0)\right|+\max\{|\mu|,|\nu|\}\leq|\kappa|+|j|
\end{equation*}
which gives the estimate \eqref{Coeff2}.
\end{proof}

According to Lemma \ref{Zeros1}, for fixed parameters $(\mu,\nu)\in\C^2$ the eigenvalues of $A(\kappa;\mu,\nu)$ are
exactly the zeros of the function $\Delta(\lambda)$ given by \eqref{Delta}. In principle, this result can be used for
numerical computation of the eigenvalues. However, in order to calculate $\Delta(\lambda)$ at some point $\lambda\in\C$,
we first have to determine the coefficients $h_n(\lambda)$ with the help of the recurrence relation \eqref{Recur} and
subsequently we need to evaluate $h(x,\lambda)$ at $x=\frac{1}{2}$ by means of the power series expansion \eqref{Funct}.
Unfortunately, this method requires the calculation of two consecutive limits, making things rather complicated. In the
remaining part of this section we show that there is yet another function $\Theta$ which encodes the eigenvalues of
$A(\kappa;\mu,\nu)$. The main advantage of $\Theta$ is, that it can be obtained by only one limit process.

By setting $y(x) := x^\alpha (1-x)^{1-\alpha}\hat y(x)$ with $\alpha := \frac{\kappa}{2}+\frac{1}{4}$, the system
\eqref{AngSys} becomes
\begin{equation} \label{MovSys}
\hat y'(x) = \left[\frac{1}{x}\,\hat B_0 + \frac{1}{x-1}\,\hat B_1 + C\right]\,\hat y(x)
\end{equation}
with the coefficient matrices
\begin{equation*}
\hat B_0 := \left(\begin{array}{cc} -\kappa-\frac{1}{2} & \mu-\lambda \\[1ex] 0 & 0  \end{array}\right),\quad
\hat B_1 := \left(\begin{array}{cc}  \kappa-\frac{1}{2} & 0 \\[1ex] \mu-\lambda & -1 \end{array}\right),\quad
 C  = \left(\begin{array}{cc} -2\nu & -2\mu \\[1ex] 2\mu & 2\nu \end{array}\right).
\end{equation*}
Now, there exists a holomorphic solution of \eqref{MovSys} in $\mathfrak{B}_1$ given by
\begin{equation}
\hat y(x,\lambda) = \sum_{n=0}^\infty x^n d_n(\lambda),\quad d_0(\lambda) := \left(\begin{array}{c} \mu-\lambda \\[1ex] \kappa+\frac{1}{2} \end{array}\right),
\end{equation}
where $d_0(\lambda)$ is an eigenvector of $\hat B_0$ for the eigenvalue $0$. In addition, the coefficients
$d_n(\lambda)$, $n>1$, are uniquely determined by the recurrence relation
\begin{equation*}
d_n(\lambda) = (\hat B_0-n)^{-1}\left[(E-n)d_{n-1}(\lambda) + C\,d_{n-2}(\lambda)\right]
\end{equation*}
with
\begin{equation*}
E := \left(\begin{array}{cc} 2\nu & 3\mu-\lambda \\[1ex] -\mu-\lambda & -2\nu \end{array}\right),\quad
d_{-1}(\lambda) := 0.
\end{equation*}
Finally, we denote by $\Theta_n(\lambda)$ the second component of $d_n(\lambda)$.

\begin{lemma} \label{Zeros2}
Let $\kappa\in[\frac{1}{2},\infty)$ and $(\mu,\nu)\in\C^2$ be fixed. Then, for each $\lambda\in\C$, the limit
\begin{equation} \label{Theta}
\Theta(\lambda) := \lim_{n\to\infty}\Theta_n(\lambda)
\end{equation}
exists, and $\Theta:\C\longrightarrow\C$ is a holomorphic function. Moreover, a point $\lambda\in\C$ is an eigenvalue
of $A(\kappa;\mu,\nu)$ if and only if $\Theta(\lambda)=0$.
\end{lemma}

\begin{proof}
For fixed $\lambda\in\C$, the differential equation \eqref{MovSys} has a regular singular point at $x=1$ with
characteristic values $-1$ and $\kappa-\frac{1}{2}$. First, let us assume that their difference $\kappa+\frac{1}{2}$
is not an integer. In this case the system \eqref{MovSys} has a fundamental system of solutions
in a complex neighbourhood of $x=1$, which can be written as
\begin{equation} \label{FundSys}
\hat y_1(x,\lambda) = (1-x)^{-1}\sum_{n=0}^\infty(1-x)^n d_n^1(\lambda),\quad
\hat y_2(x,\lambda) = (1-x)^{\kappa-\frac{1}{2}}\sum_{n=0}^\infty(1-x)^n d_n^2(\lambda),
\end{equation}
where
\begin{equation*}
d_0^1(\lambda) = \left(\begin{array}{cc}  0 \\[1ex] 1 \end{array}\right) =: e_2,\quad
d_0^2(\lambda) = \left(\begin{array}{cc}  \kappa+\frac{1}{2} \\[1ex] \mu-\lambda \end{array}\right)
\end{equation*}
are eigenvectors of $\hat B_1$ for the eigenvalues $-1$ and $\kappa-\frac{1}{2}$, respectively. Now, $\hat y$ can be
written as a linear combination
\begin{equation*}
\hat y(x,\lambda) = \gamma_1(\lambda)\hat y_1(x,\lambda) + \gamma_2(\lambda)\hat y_2(x,\lambda)
\end{equation*}
with connection coefficients $\gamma_1(\lambda),\,\gamma_2(\lambda)\in\C$. Applying \cite[Corollary 1.6]{SS}
to the system \eqref{MovSys} gives
\begin{equation} \label{gamma}
\lim_{n\to\infty}d_n(\lambda) = \gamma_1(\lambda)e_2,
\end{equation}
and therefore the limit \eqref{Theta} exists. Furthermore, $\lambda$ is an eigenvalue of $A(\kappa;\mu,\nu)$ if and only
if $\gamma_1(\lambda)=0$, i.e., if and only if $\Theta(\lambda)$ becomes zero. Finally, it can be shown that the functions
$d_n$ converge uniformly in every compact subset of $\C$, and Weierstrass' theorem implies that
$\Theta$ is an entire function.

Now, suppose that $k:=\kappa+\frac{1}{2}$ is a positive integer. In this case, a fundamental system of the form
\eqref{FundSys} may not exist. Nevertheless, it can be proved (see Lemma \ref{Fundamental} in Section VI) that the
system \eqref{MovSys} has a fundamental matrix
\begin{equation*}
\hat Y(x,\lambda) = G(\lambda)\sum_{n=0}^\infty H_n(\lambda)(1-x)^n(1-x)^D(1-x)^{J(\lambda)},
\end{equation*}
in a complex neighbourhood of $x=1$, where $D := \diag(-1,k-1)$, $H_0(\lambda)=I$ and
\begin{equation*}
G(\lambda) = \left(\begin{array}{cc} 0  & \kappa+\frac{1}{2} \\[1ex] 1 & \mu-\lambda \end{array}\right),\quad
J(\lambda) = \left(\begin{array}{cc} 0 & 0 \\[1ex] q(\lambda) & 0 \end{array}\right)
\end{equation*}
with some $q(\lambda)\in\C$. In particular, we can write $\hat Y$ in the form
\begin{equation*}
\hat Y(x,\lambda) = \hat H(x,\lambda)(1-x)^{\tilde J(\lambda)},\quad \hat H(x,\lambda)=\sum_{n=0}^\infty(1-x)^n D_n(\lambda),
\end{equation*}
where
\begin{equation*}
D_0(\lambda) = \left(\begin{array}{cc} 0  & 0 \\[1ex] 1 & 0 \end{array}\right),\quad \tilde J(\lambda) = \left(\begin{array}{cc} -1 & 0 \\[1ex] q(\lambda) & -1 \end{array}\right)
\end{equation*}
Since $\hat y$ solves the system \eqref{MovSys}, there exists a vector $c(\lambda)\in\C^2$ such that
$\hat y(x,\lambda)=\hat Y(x,\lambda)c(\lambda)$, and \cite[Theorem 1.1]{SchaefkeR} implies
\begin{equation} \label{Gamma}
d_n(\lambda) = D_0(\lambda)\frac{1}{\Gamma}\left(-\tilde J(\lambda)\right)\Gamma(n+1)\frac{1}{\Gamma}\left(n-\tilde J(\lambda)\right)c(\lambda) + \Osym\left(n^{\delta-1}\right)
\end{equation}
for arbitrary $\delta>0$. For the definition and discussion of the reciprocal Gamma function for matrices we refer to the
Appendix in \cite{SchaefkeR}. Particularly, for the Jordan type matrices $-\tilde J(\lambda)$ and $n-\tilde J(\lambda)$ i
we obtain
\begin{equation*}
\frac{1}{\Gamma}\left(-\tilde J(\lambda)\right)  = \left(\begin{array}{cc} 1 & 0 \\[1ex] \ast & 1 \end{array}\right),\quad
\frac{1}{\Gamma}\left(n-\tilde J(\lambda)\right) = \left(\begin{array}{cc} \frac{1}{\Gamma(n+1)} & 0 \\[1ex] \ast & \frac{1}{\Gamma(n+1)} \end{array}\right).
\end{equation*}
Now, if $\gamma_1(\lambda)$ denotes the first component of $c(\lambda)$, then \eqref{Gamma} implies \eqref{gamma}. Since
$\lambda$ is an eigenvalue of $A(\kappa;\mu,\nu)$ if and only if $\gamma_1(\lambda)=0$, the proof of Lemma \ref{Zeros2}
is complete.
\end{proof}

\section{A partial differential equation for the eigenvalues}

\begin{theorem} \label{KPT}
For fixed $\kappa\in[\frac{1}{2},\infty)$ and $j\in\Z\setminus\{0\}$, the $j$-th eigenvalue
$\lambda=\lambda_j(\kappa;\mu,\nu)$ of $A$ is an analytical function in $(\mu,\nu)\in\R^2$ satisfying
the first order quasi-linear partial differential equation
\begin{equation} \label{PDE}
\left(\mu-2\,\nu\,\lambda\right)\frac{\partial\lambda}{\partial\mu} +
\left(\nu-2\,\mu\,\lambda\right)\frac{\partial\lambda}{\partial\nu} + 2\,\kappa\,\mu + 2\,\mu\,\nu = 0,
\end{equation}
where $\lambda_j(\kappa;0,0)$ is given by \eqref{Init}.
\end{theorem}

\begin{proof}
Let
\begin{equation*}
S(\theta) =: \left(\begin{array}{c} S_1(\theta) \\[1ex] S_2(\theta) \end{array}\right),\quad \theta\in(0,\pi),
\end{equation*}
be that eigenfunction of $A(\kappa;\mu,\nu)$ for the eigenvalue $\lambda=\lambda_{j}(\kappa;\mu,\nu)$ which is normalised
by the condition $(S,S)=1$. Introducing the functions
\begin{equation*}
U(\theta) := S_1(\theta)^2+S_2(\theta)^2,\quad
V(\theta) := S_2(\theta)^2-S_1(\theta)^2,\quad
W(\theta) := 2\,S_1(\theta)\,S_2(\theta),
\end{equation*}
a straightforward calculation shows that $U$, $V$, and $W$ are solutions of the system of differential equations
\begin{align}
U'(\theta) & = 2\left(\nu\sin\theta+\frac{\kappa}{\sin\theta}\right)V(\theta)+2\,\mu\cos\theta\,W(\theta), \label{Der1} \\
V'(\theta) & = 2\left(\nu\sin\theta+\frac{\kappa}{\sin\theta}\right)U(\theta)+2\,\lambda\,W(\theta), \label{Der2} \\
W'(\theta) & = 2\,\mu\cos\theta\,U(\theta)-2\,\lambda\,V(\theta). \label{Der3}
\end{align}
Now, from analytic perturbation theory (compare \cite[Chap. VII, \S 3, Sec. 4]{Kato}) it follows that
\begin{align}
\frac{\partial \lambda}{\partial\mu}
& = \left(\frac{\partial A}{\partial\mu}\,S,S\right)
  = \int_0^{\pi}S(\theta)^{\ast}\left(\begin{array}{cc} -\cos\theta & 0 \\[1ex] 0 & \cos\theta \end{array}\right)S(\theta)\,\d\theta
  = \int_0^{\pi}\cos\theta\,V(\theta)\,\d\theta, \label{Par1} \\
\frac{\partial \lambda}{\partial\nu}
& = \left(\frac{\partial A}{\partial\nu}\,S,S\right)
  = \int_0^{\pi}S(\theta)^{\ast}\left(\begin{array}{cc} 0 & -\sin\theta \\[1ex] -\sin\theta & 0 \end{array}\right)S(\theta)\,\d\theta
  = -\int_0^{\pi}\sin\theta\,W(\theta)\,\d\theta. \label{Par2}
\end{align}
In addition, from \eqref{EigEst} we obtain the estimates
\begin{equation*}
|U(\theta)|,\,|V(\theta)|,\,|W(\theta)|\leq C\sin^{2\kappa}\theta
\end{equation*}
with some constant $C>0$. Since $\kappa$ is positive, $U$, $V$ and $W$ vanish at $\theta=0$
and $\theta=\pi$. If we integrate \eqref{Par1} by parts and replace $V'(\theta)$ with the r.h.s. of \eqref{Der2}, then
we get
\begin{align*}
\frac{\partial \lambda}{\partial\mu}
& = -\int_0^{\pi}\sin\theta\,V'(\theta)\,\d\theta
  = -\int_0^{\pi}\left(2\,\nu\sin^2\theta+2\,\kappa\right)U(\theta)+2\,\lambda\sin\theta\,W(\theta)\,\d\theta \notag \\
& = -(2\,\nu + 2\,\kappa)\int_0^{\pi}U(\theta)\,\d\theta - 2\,\lambda\int_0^{\pi}\sin\theta\,W(\theta)\,\d\theta + 2\,\nu\int_0^{\pi}\cos^2\theta\,U(\theta)\,\d\theta.
\end{align*}
Taking into account that
\begin{equation*}
\int_0^{\pi}U(\theta)\,\d\theta = (S,S) = 1,\quad\int_0^{\pi}\sin\theta\,W(\theta)\,\d\theta = -\frac{\partial \lambda}{\partial\nu},
\end{equation*}
we have
\begin{equation} \label{Equ}
\mu\,\frac{\partial \lambda}{\partial\mu}
 = -\mu\,(2\,\nu + 2\,\kappa) + 2\,\mu\,\lambda\,\frac{\partial \lambda}{\partial\nu} + 2\,\mu\,\nu\int_0^{\pi}\cos^2\theta\,U(\theta)\,\d\theta.
\end{equation}
Moreover, equation \eqref{Der3} implies
\begin{equation*}
2\,\mu\cos^2\theta\,U(\theta) = \cos\theta\,W'(\theta)+2\,\lambda\cos\theta\,V(\theta),
\end{equation*}
and integration by parts gives
\begin{align}
2\,\mu\,\nu\int_0^{\pi}\cos^2\theta\,U(\theta)\,\d\theta
& = \nu\int_{0}^{\pi}\cos\theta\,W'(\theta)\,\d\theta + 2\,\nu\,\lambda\int_{0}^{\pi}\cos\theta\,V(\theta)\,\d\theta \notag \\
& = \nu\int_{0}^{\pi}\sin\theta\,W(\theta)\,\d\theta + 2\,\nu\,\lambda\int_{0}^{\pi}\cos\theta\,V(\theta)\,\d\theta
  = -\nu\,\frac{\partial\lambda}{\partial\nu} + 2\,\nu\,\lambda\,\frac{\partial\lambda}{\partial\mu}. \label{Int}
\end{align}
Replacing the last term on the r.h.s. of \eqref{Equ} with \eqref{Int}, we obtain exactly the partial differential equation
\eqref{PDE}.
\end{proof}

The PDE \eqref{PDE} can be used in order to derive a power series expansion for $\lambda_j$ with respect to $\mu$ and
$\nu$. For this purpose we introduce the new coordinates (compare \cite{SFC})
\begin{equation*}
\alpha := \nu-\mu,\quad \beta := \nu+\mu.
\end{equation*}
Then $\hat\lambda(\alpha,\beta):=\lambda_j\left(\kappa;\frac{\beta-\alpha}{2},\frac{\beta+\alpha}{2}\right)$ is a solution
of the transformed partial differential equation
\begin{equation} \label{PDE1}
\alpha\left(1+2\,\hat\lambda\right)\frac{\partial\hat\lambda}{\partial\alpha} +
\beta \left(1-2\,\hat\lambda\right)\frac{\partial\hat\lambda}{\partial\beta}  = \kappa(\alpha-\beta)+\frac{1}{2}\left(\alpha^2-\beta^2\right),
\end{equation}
where $\hat\lambda(0,0)=\lambda_j(\kappa;0,0)$ is given by \eqref{Init}. As $\hat\lambda$ depends analytically on
$(\alpha,\beta)$, there exists a series expansion for $\hat\lambda$ of the form
\begin{equation} \label{Series}
\hat\lambda(\alpha,\beta) = \sum_{m,n=0}^{\infty}c_{m,n}\alpha^m\beta^n
\end{equation}
(for clarity, the indices $\kappa$ and $j$ in the coefficients $c_{m,n}$ and in the function $\hat\lambda$ have been
omitted). Furthermore, \eqref{PDE1} is equivalent to
\begin{equation} \label{PDE2}
\alpha\left(\frac{\partial\hat\lambda}{\partial\alpha} + \frac{\partial\hat\lambda^2}{\partial\alpha}\right) +
\beta \left(\frac{\partial\hat\lambda}{\partial\beta}  - \frac{\partial\hat\lambda^2}{\partial\beta} \right) = \kappa(\alpha-\beta)+\frac{1}{2}\left(\alpha^2-\beta^2\right),
\end{equation}
and since
\begin{equation*}
\hat\lambda(\alpha,\beta)^2 = \sum_{m,n=0}^{\infty}\left(\sum_{r=0}^m\sum_{s=0}^n c_{r,s}\,c_{m-r,n-s}\right)\alpha^m\beta^n,
\end{equation*}
we obtain the identity
\begin{equation*} \label{Pow}
\sum_{m,n=0}^{\infty}\left((m+n)c_{m,n}+(m-n)\sum_{r=0}^m\sum_{s=0}^n c_{r,s}\,c_{m-r,n-s}\right)\alpha^m\beta^n = \kappa(\alpha-\beta)+\frac{1}{2}\left(\alpha^2-\beta^2\right).
\end{equation*}
Comparing the terms of equal order in $\alpha$ and $\beta$, it follows that
\begin{gather*}
c_{0,0} = \lambda_j(\kappa;0,0) =: c_0,\quad c_{1,0} = \frac{\kappa}{2\,c_0+1},\quad c_{0,1} = \frac{\kappa}{2\,c_0-1}, \\
c_{2,0} = \frac{(2\,c_0+1)^2-4\,\kappa^2}{4(2\,c_0+1)^3},\quad c_{1,1} = 0,\quad c_{0,2} = \frac{(2\,c_0-1)^2-4\,\kappa^2}{4(2\,c_0-1)^3},
\end{gather*}
and for $m+n>2$ the coefficients $c_{m,n}$ satisfy
\begin{equation} \label{Rec1}
\left((m+n)+2\,c_0(m-n)\right)c_{m,n} = (n-m)\sum_{(r,s)\in\llbracket m,n\rrbracket}c_{r,s}\,c_{m-r,n-s}
\end{equation}
where $\llbracket m,n\rrbracket$ denotes the set of all pairs $(r,s)\in\Z^2$ with $0\leq r\leq m$, $0\leq s\leq n$ and
$0<r+s<m+n$. In particular, if $m=n>0$, then we get $2\,n\,c_{n,n}=0$, which implies
\begin{equation} \label{Diag}
c_{n,n}=0\qquad\mbox{for all}\quad n>0.
\end{equation}
Moreover, if $\kappa$ is not a rational number, i.e., $\kappa\in[\frac{1}{2},\infty)\setminus\Q$, then the initial value
$c_0$ is not a rational number, and we have $(m+n)+2\,c_0(m-n)\neq 0$ for all $(m,n)\in\Z^2$ with $m+n>2$. In this case
\eqref{Rec1} gives a recurrence formula for all coefficients of the power series expansion \eqref{Series}.

Now, suppose that $\kappa\in[\frac{1}{2},\infty)\cap\Q$. Then $c_0$ is a rational number with $|c_0|\geq 1$, and we get
$\frac{2\,c_0-1}{2\,c_0+1}=\frac{p}{q}$ with some coprime integers $p$ and $q$. Now, the prefactor on the l.h.s. of
\eqref{Rec1} becomes zero if and only if $m=\ell p$, $n=\ell q$ with some positive integer $\ell$, and thus the
coefficients $c_{\ell p,\ell q}$ are not determined by \eqref{Rec1}. However, we can by-pass this problem if we regard
$\kappa$ as an additional parameter in our eigenvalue problem. Since the coefficient matrix of \eqref{AngSys} depends
holomorphically on $\kappa\in\C^{+} := \{z\in\C:\re z>0\}$ and $(\lambda,\mu,\nu)\in\C^3$, we obtain in a similar way as
described in Section 1 that $h$ in \eqref{Holom} and therefore $\Delta=\Delta(\kappa;\lambda;\mu,\nu)$ given by
\eqref{Delta} is a holomorphic function on $\C^{+}\times\C^3$. Moreover, in the same way as in the proof of Lemma
\ref{Zeros1}, we can show that for fixed $\kappa\in[\frac{1}{2},\infty)$ and $(\mu,\nu)\in\C^2$ the eigenvalues of
$A(\kappa;\mu,\nu)$ coincide with the zeros of the function $\lambda\longmapsto\Delta(\kappa;\lambda;\mu,\nu)$. In the
case $(\kappa,\mu,\nu)\in[\frac{1}{2},\infty)\times\R^2$ these zeros are simple, since $A(\kappa;\mu,\nu)$ has only
simple eigenvalues. Hence, by solving the equation $\Delta(\kappa;\lambda;\mu,\nu)=0$, we find that an eigenvalue
$\lambda_j(\kappa;\mu,\nu)$ is a holomorphic function in a complex neighbourhood of $[\frac{1}{2},\infty)\times\R^2$. In
particular, $\hat\lambda$ depends holomorphically on $(\kappa;\alpha,\beta)$, and for a given
$\kappa\in[\frac{1}{2},\infty)$, there exists a power series expansion of the form
\begin{equation*}
\hat\lambda(\kappa+\veps;\alpha,\beta) = \sum_{l,m,n=0}^{\infty}c_{m,n}^{(l)}\veps^l\alpha^m\beta^n
\end{equation*}
in a neighbourhood of $(\kappa,0,0)$. In the following we derive a recurrence relation for the coefficients
$c_{m,n}^{(l)}$. Since
\begin{equation*}
\hat\lambda(\kappa+\veps;\alpha,\beta)^2 = \sum_{l,m,n=0}^{\infty}
\left(\sum_{t=0}^l\sum_{r=0}^m\sum_{s=0}^n c_{r,s}^{(t)}\,c_{m-r,n-s}^{(l-t)}\right)\veps^l\alpha^m\beta^n,
\end{equation*}
from \eqref{PDE2} it follows that
\begin{equation} \label{Sum}
\sum_{l,m,n=0}^{\infty}\left((m+n)c_{m,n}^{(l)}+(m-n)\sum_{t=0}^l\sum_{r=0}^m\sum_{s=0}^n
c_{r,s}^{(t)}\,c_{m-r,n-s}^{(l-t)}\right)\veps^l\alpha^m\beta^n
\ =\ \kappa(\alpha-\beta) + \veps(\alpha-\beta) + \frac{1}{2}\left(\alpha^2-\beta^2\right).
\end{equation}
Moreover, \eqref{Init} implies that
\begin{equation*}
c_{0,0}^{(l)} = \frac{1}{l!}\,\frac{\partial^l\lambda}{\partial\kappa^l}(\kappa;0,0) =
\left\{\begin{array}{rl}
\sign(j)(\kappa-\frac{1}{2}+|j|), & \mbox{ if }l=0, \\[1ex]
\sign(j), & \mbox{ if }l=1,\\[1ex]
0, & \mbox{ if }l>1.\end{array}\right.
\end{equation*}
Comparing the terms of equal order in \eqref{Sum}, we obtain
\begin{gather*}
c_{0,0}^{(0)} = \lambda_j(\kappa;0,0) =: c_0,\quad
c_{1,0}^{(0)} = \frac{\kappa}{2\,c_0+1},\quad
c_{0,1}^{(0)} = \frac{\kappa}{2\,c_0-1},\\
c_{2,0}^{(0)} = \frac{(2\,c_0+1)^2-4\,\kappa^2}{4(2\,c_0+1)^3},\quad
c_{1,1}^{(0)} = 0,\quad
c_{0,2}^{(0)} = \frac{(2\,c_0-1)^2-4\,\kappa^2}{4(2\,c_0-1)^3}\\
c_{1,0}^{(1)} = \frac{2\,c_0+1-2\,\sign(j)\,\kappa}{(2\,c_0+1)^2},\quad
c_{0,1}^{(1)} = \frac{2\,c_0-1-2\,\sign(j)\,\kappa}{(2\,c_0-1)^2},
\end{gather*}
while the remaining coefficients are determined by the identity
\begin{equation} \label{Rec2}
\left((m+n)+2\,c_0(m-n)\right)c_{m,n}^{(l)} + (m-n)\sum_{(t,r,s)\in\llbracket l,m,n\rrbracket}c_{r,s}^{(t)}\,c_{m-r,n-s}^{(l-t)} = 0,\quad l+m+n>2.
\end{equation}
Here $\llbracket l,m,n\rrbracket$ denotes the set of all triples $(t,r,s)\in\Z^3$ with $0\leq t\leq l$,
$0\leq r\leq m$, $0\leq s\leq n$, and $0<t+r+s<l+m+n$. In the case $(m+n)+2\,c_0(m-n)=0$, the prefactor of
$c_{m,n}^{(l)}$ in \eqref{Rec2} vanishes, and since $m-n\neq 0$, we get for $l>0$
\begin{equation} \label{Rec3}
0 = \sum_{(t,r,s)\in\llbracket l,m,n\rrbracket}c_{r,s}^{(t)}\,c_{m-r,n-s}^{(l-t)}
  = 2\,c_{0,0}^{(1)}\,c_{m,n}^{(l-1)} + \sum_{(t,r,s)\in\llbracket l,m,n\rrbracket^\ast}c_{r,s}^{(t)}\,c_{m-r,n-s}^{(l-t)},
\end{equation}
where $\llbracket l,m,n\rrbracket^\ast := \llbracket l,m,n\rrbracket\setminus\left\{(1,0,0),(l-1,m,n)\right\}$.
Now, for all coefficients $c_{m,n}^{(l)}$ with $l+m+n>2$, \eqref{Rec2} implies
\begin{equation*}
c_{m,n}^{(l)} = \frac{n-m}{(m+n)+2\,c_0(m-n)}\sum_{(t,r,s)\in\llbracket l,m,n\rrbracket}c_{r,s}^{(t)}\,c_{m-r,n-s}^{(l-t)}\qquad
\mbox{if}\quad (m+n)+2\,c_0(m-n)\neq 0,
\end{equation*}
whereas \eqref{Rec3} and $c_{0,0}^{(1)}=\sign(j)$ yield
\begin{equation*}
c_{m,n}^{(l-1)} = -\frac{\sign(j)}{2}\sum_{(t,r,s)\in\llbracket l,m,n\rrbracket^\ast}c_{r,s}^{(t)}\,c_{m-r,n-s}^{(l-t)}\qquad
\mbox{if }\quad (m+n)+2\,c_0(m-n) = 0\mbox{ and }l > 1.
\end{equation*}
These recurrence relations can be used to determine all the coefficients $c_{m,n} = c_{m,n}^{(0)}$ of the power
series expansion \eqref{Series} in the case that $\kappa$ is a rational number.

\begin{remark}
A series expansion for the eigenvalues $\hat\lambda$ with respect to $(\alpha,\beta)$ has been given by Suffern,
Fackerell \& Cosgrove \cite[Sec. 8]{SFC}, however, only the coefficients $c_{m,n}$ with $m+n\leq 5$ have been
determined. Furthermore, Kalnins \& Miller \cite{KM} studied a series expansion $\lambda=\sum_{n=0}^\infty\lambda_n a^n$
for the eigenvalues in terms of the Kerr parameter $a$, but also in this paper only a finite number of coefficients
$\lambda_0,\ldots,\lambda_3$ have been explicitly computed. A general recurrence relation for the coefficients of
\eqref{Series} could not be found in the literature. Moreover, the problem of dividing by numbers which may be zero has
not been noticed in \cite{SFC} and \cite{KM}. Finally, it should be noted that some of the diagonal entries $c_{n,n}$ for
$n>0$ in \cite[Table I]{SFC} are not equal to zero, in contrast to our result \eqref{Diag}.
\end{remark}

\section{Solution of the PDE by the method of characteristics}

In this section we study the PDE \eqref{PDE} for real parameters $(\mu,\nu)\in\R^2$ and fixed
$\kappa\in[\frac{1}{2},\infty)$ by the method of characteristics. In particular, we obtain an exact formula for the
eigenvalues in the case $|\mu|=|\nu|$, and for $|\mu|\neq|\nu|$, it turns out that the characteristic equations can be
reduced to the third Painlev\'e equation.

\begin{theorem} \label{Formula}
Let $\kappa\in[\frac{1}{2},\infty)$, $j\in\Z\setminus\{0\}$ and $\tau\in\{-1,+1\}$ be fixed. Then
\begin{equation} \label{Sol}
\lambda_j(\kappa;\mu,\tau\mu) = \frac{\tau}{2}+\sign(j)\sqrt{\left(\lambda_j(\kappa;0,0)-\frac{\tau}{2}\right)^2 + 2\,\tau\kappa\mu + \mu^2},
\end{equation}
where $\lambda_j(\kappa;0,0)=\sign(j)\left(\kappa-\frac{1}{2}+|j|\right)$. In particular, if $j=\tau$, then
\begin{equation*}
\lambda_j(\kappa;\mu,\tau\mu) = \tau\left(\kappa+\frac{1}{2}\right)+\mu.
\end{equation*}
\end{theorem}

\begin{proof}
According to Theorem \ref{KPT}, the function $\lambda(\mu,\nu):=\lambda_j(\kappa;\mu,\nu)$ solves the partial differential
equation \eqref{PDE}. Defining $w(\mu):=\lambda\left(\mu,\tau\mu\right)$, $\mu\in\R$, for some fixed $\tau\in\{-1,+1\}$,
we obtain
\begin{equation*}
w'(\mu) = \frac{\partial\lambda}{\partial\mu}\left(\mu,\tau\mu\right) + \tau\frac{\partial\lambda}{\partial\nu}\left(\mu,\tau\mu\right),
\end{equation*}
and with the help of \eqref{PDE} it can be shown that
\begin{equation*}
\mu\,w'(\mu) - 2\,\tau\mu\,w(\mu)w'(\mu) = -2\,\kappa\mu - 2\,\tau \mu^2.
\end{equation*}
Dividing the above differential equation by $-\tau\mu$ and integrating gives
\begin{equation} \label{Form1}
\left(w(\mu)-\frac{\tau}{2}\right)^2 = C + 2\,\tau\kappa\mu + \mu^2,\quad \mu\in\R,
\end{equation}
where the constant of integration $C$ is uniquely determined by
\begin{equation*}
C = \left(w(0)-\frac{\tau}{2}\right)^2 = \left(\lambda_j(\kappa;0,0)-\frac{\tau}{2}\right)^2.
\end{equation*}
Now, from \eqref{Form1} it follows that
\begin{equation} \label{Form2}
w(\mu) = \frac{\tau}{2}+\veps\,\sqrt{\left(\lambda_j(\kappa;0,0)-\frac{\tau}{2}\right)^2+2\,\tau\kappa\mu + \mu^2}
\end{equation}
with some $\veps\in\{-1,+1\}$ and the square root assumed to be non-negative. We have to take the sign $\veps$ such
that the l.h.s. of \eqref{Form2} is analytic and coincides with $\lambda_j(\kappa;0,0)$ at the point $\mu=0$. If $j=\tau$,
then $\lambda_j(\kappa;0,0)=\tau\left(\kappa+\frac{1}{2}\right)$, and \eqref{Form2} implies
$w(\mu) = \frac{\tau}{2}+\veps\left(\tau\kappa+\mu\right)$. Inserting $\mu=0$, it follows that $\veps=\tau$, i.e.,
$w(\mu) = \tau\left(\kappa+\frac{1}{2}\right)+\mu$. In the case $j\neq\tau$ we have
$\left|\lambda_j(\kappa;0,0)-\frac{\tau}{2}\right|\geq \kappa+1$ and thus the radicand in \eqref{Form2} is positive
for all $\mu\in\R$. Moreover, by means of
\begin{equation*}
\lambda_j(\kappa;0,0) = w(0) = \frac{\tau}{2}+\veps\,\sqrt{\left(\lambda_j(\kappa;0,0)-\frac{\tau}{2}\right)^2}
= \frac{\tau}{2}+\veps\left|\lambda_j(\kappa;0,0)-\frac{\tau}{2}\right|
\end{equation*}
and \eqref{Init}, we get $\veps = \sign\left(\lambda_j(\kappa;0,0)-\frac{\tau}{2}\right) = \sign(j)$, which completes
the proof.
\end{proof}

\begin{remark}
For a given half-integer $\kappa$ and $\mu=\nu$, this result has been shown by Suffern, Fackerell \& Cosgrove using a
power series expansion for the eigenfunctions of \eqref{CPE1} -- \eqref{CPE2} in terms of hypergeometric functions (see
\cite[Sec. 3--5]{SFC}). Here, we obtained the formula for $\lambda_j(\kappa;\mu,\pm\mu)$ as an immediate consequence
of the partial differential equation \eqref{PDE}. Moreover, it should be noted that the formula \cite[(54)]{Chakrabarti}
given by Chakrabarti is not correct.
\end{remark}

Now, let us consider the case $|\mu|\neq|\nu|$. For this reason, we introduce new coordinates
$(t,v)\in(0,\infty)\times\left(\R\setminus\{0\}\right)$ by
\begin{equation} \label{tv}
\mu(t,v) = \frac{t}{2}\left(v+\frac{\sigma}{v}\right),\quad\nu(t,v) = \frac{t}{2}\left(v-\frac{\sigma}{v}\right)
\end{equation}
with some fixed $\sigma\in\{-1,+1\}$ (note that $\sigma=\pm 1$ corresponds to the cases $|\mu|>|\nu|$ and $|\mu|<|\nu|$,
respectively; moreover, this transformation maps $v=\const.$ onto lines in the $(\mu,\nu)$-plane starting at the origin,
while the curves $t=\const.$ are mapped onto hyperboles). By setting $w(t,v)=\lambda(\mu,\nu)$, we have
\begin{align*}
\frac{\partial w}{\partial t}
& = \frac{1}{2}\left(v+\frac{\sigma}{v}\right)\frac{\partial\lambda}{\partial\mu}+\frac{1}{2}\left(v-\frac{\sigma}{v}\right)\frac{\partial\lambda}{\partial\nu}
  = \frac{1}{t}\left(\mu\,\frac{\partial\lambda}{\partial\mu}+\nu\,\frac{\partial\lambda}{\partial\nu}\right), \\
\frac{\partial w}{\partial v}
& = \frac{t}{2}\left(1-\frac{\sigma}{v^2}\right)\frac{\partial\lambda}{\partial\mu}+\frac{t}{2}\left(1+\frac{\sigma}{v^2}\right)\frac{\partial\lambda}{\partial\nu}
  = \frac{1}{v}\left(\nu\,\frac{\partial\lambda}{\partial\mu}+\mu\,\frac{\partial\lambda}{\partial\nu}\right),
\end{align*}
and \eqref{PDE} becomes
\begin{equation} \label{PDEw}
\frac{\partial w}{\partial t}-\frac{2\,v\,w}{t}\frac{\partial w}{\partial v} + \kappa\left(v+\frac{\sigma}{v}\right) + \frac{t}{2}\left(v^2-\frac{1}{v^2}\right) = 0.
\end{equation}
The characteristic equations of this PDE are given by
\begin{align}
v'(t) & = -\frac{2\,v(t)\,w(t)}{t}, \label{Char1} \\
w'(t) & = -\kappa\left(v(t)+\frac{\sigma}{v(t)}\right) - \frac{t}{2}\left(v(t)^2-\frac{1}{v(t)^2}\right). \label{Char2}
\end{align}
From \eqref{Char1} we obtain that $w(t) = -\frac{t\,v'(t)}{2\,v(t)}$, and \eqref{Char2} implies
\begin{equation*}
\frac{v'(t)}{2\,v(t)} + \frac{t\,v''(t)}{2\,v(t)} - \frac{t\,v'(t)^2}{2\,v(t)^2}
= \kappa\left(v(t)+\frac{\sigma}{v(t)}\right) + \frac{t}{2}\left(v(t)^2-\frac{1}{v(t)^2}\right).
\end{equation*}
Multiplying the above differential equation with $2\,v(t)^2$, we get the following third Painlev\'e equation
\begin{equation} \label{PIII}
t\,v\,v'' - t(v')^2 + v\,v' - 2\kappa\left(v^2 + \sigma\right)v - t\left(v^4-1\right) = 0,
\end{equation}
with parameters $\alpha=\sigma\beta=2\kappa$ and $\gamma=-\delta=1$ (see, for example, \cite{MCB} or \cite{MW}). For
further details on the Painlev{\'e} III we refer to e.g. \cite{CT}, \cite{Widom} and \cite{IKSY}.

In general, Painlev{\'e} III is not solvable in terms of elementary functions, and therefore we cannot expect a closed
expression for the eigenvalues of $A(\kappa;\mu,\nu)$ in the case $|\mu|\neq|\nu|$. On the other hand, for particular
values of $\kappa$ there exist so-called special integrals of polynomial type for this equation, i.e., polynomials $Q$
in $t$, $v$ and $v'$ with the property that every solution of the differential equation $Q(t,v,v')=0$ satisfies
\eqref{PIII}. As it will be shown below, such special integrals are related to algebraic solutions of the PDE
\eqref{PDE}, i.e., solutions, which are zeros of a polynomial in $\lambda$
with rational coefficients in $\mu$ and $\nu$. Moreover, taking into account that the eigenvalues
$\lambda_j(\kappa;\mu,\tau\mu)$, $\tau\in\{-1,1\}$, of $A(\kappa;\mu,\tau\mu)$ satisfy the quadratic equation
\begin{equation*}
\left(\lambda-\frac{\tau}{2}\right)^2 = C + 2\,\tau\kappa\,\mu + \mu^2\quad\mbox{ with }\quad C := \left(\lambda_j(\kappa;0,0)-\frac{\tau}{2}\right)^2,
\end{equation*}
the question arises if such an algebraic expression for the eigenvalues of
$A(\kappa;\mu,\nu)$ exists in the case $|\mu|\neq|\nu|$. A first step towards the answer of this problem is given by the
next Lemma.

\begin{lemma} \label{AlgSol}
Suppose that there exists a polynomial
\begin{equation*}
P(\lambda;\mu,\nu) = \sum_{n=0}^N P_n(\mu,\nu)\lambda^n,\quad P_N \equiv 1,
\end{equation*}
of degree $N>0$ in $\lambda$ with rational coefficients $P_n$ in $\mu$ and $\nu$ such that the zeros
$z_j(\mu,\nu)$, $j=1,\ldots N$, of $P(\sdot;\mu,\nu)$ are simple, and that the functions $\lambda=z_j$ are
solutions of the partial differential equation \eqref{PDE}. Then $\kappa$ is a half-integer. Moreover, if
$N=1$ or $N=2$, then $\kappa=\frac{1}{2}$ and $P(\lambda;\mu,\nu)=(\lambda+\mu)^N$.
\end{lemma}

\begin{proof}
Let $Q(t,v,v') = \sum_{n=0}^N Q_n(t,v)(v')^n$ be the polynomial in $v'$ with coefficients
\begin{equation*}
Q_n(t,v) := \left(-\frac{2\,v}{t}\right)^{N-n}P_n\left(\mu(t,v),\nu(t,v)\right),\quad n=0,\ldots,N,
\end{equation*}
where $\mu(t,v)$ and $\nu(t,v)$ are given by \eqref{tv}. Note that the $Q_n$ are again rational functions
in $t$ and $v$. Moreover, let $v:\mathfrak{D}\longrightarrow\R\setminus\{0\}$ be any solution of the first order ODE
$Q(t,v,v')=0$ on some interval $\mathfrak{D}\subset\R\setminus\{0\}$. For the function
\begin{equation} \label{Diff}
w(t) = -\frac{t\,v'(t)}{2\,v(t)},\quad t\in\mathfrak{D},
\end{equation}
we obtain
\begin{equation*}
0 = Q\left(t,v(t),v'(t)\right)=\sum_{n=0}^N Q_n\left(t,v(t)\right)\left(-\frac{2\,v(t)\,w(t)}{t}\right)^n
= \left(-\frac{2\,v(t)}{t}\right)^N P\left(w(t);\mu(t,v(t)),\nu(t,v(t))\right),
\end{equation*}
and thus $w(t)$ is a zero of $P\left(\sdot;\mu\left(t,v(t)\right),\nu\left(t,v(t)\right)\right)$ for each
$t\in\mathfrak{D}$. Since the zeros of this polynomial depend analytically on the parameter $t$ according to the implicit
function theorem, there exists an index $j\in\{1,\ldots,N\}$ such that
$w(t) = z_j\left(\mu\left(t,v(t)\right),\nu\left(t,v(t)\right)\right)$ for all $t\in\mathfrak{D}$. Furthermore, as $z_j$
solves the PDE \eqref{PDE}, it follows that $(t,v(t),w(t))$, $t\in\mathfrak{D}$, is a characteristic curve of
\eqref{PDEw}, and thus $v$ is a solution of \eqref{PIII}. Hence, $Q(t,v,v')=0$ implies \eqref{PIII}, and therefore $Q$ is
a special integral of rational type for this Painlev{\'e} III. Multiplying $Q(t,v,v')$ by an appropriate polynomial
$r(t,v)$ in $t$ and $v$, we obtain that $\tilde Q(t,v,v'):=r(t,v)Q(t,v,v')$
is a special integral of polynomial type of degree $N$ with respect to $v'$. Now, \cite[Theorem 2]{MW} yields that such a
special integral exists if and only if $2\,\kappa \pm 2\,\sigma\kappa = 2(2k-1)$ with some integer $k$, i.e.,
$\kappa=k-\frac{1}{2}$ is a half-integer. In addition, by \cite[Lemma 3]{MW}, the relation $(\sigma q - p)\kappa = N$
has to be satisfied for some integers $p,\,q\in\{-N,-N+2,\ldots,N-2,N\}$. In the case $N=1$ or $N=2$, these conditions
imply $\kappa=\frac{1}{2}$, and the corresponding special integrals of polynomial type are explicitly known, namely
$r(t)\,v^s\left(v'+v^2+\sigma\right)^N$, where $r$ is some polynomial in $t$, and $s$ is an integer
(compare \cite[Section 2]{MW}). Hence, $Q(t,v,v')=\left(v'+v^2+\sigma\right)^N$ and
\begin{equation*}
P\left(w(t);\mu(t,v(t)),\nu(t,v(t))\right) = \left(-\frac{t}{2\,v(t)}\right)^N Q\left(t,v(t)\right) = \left(w(t)+\mu\left((t,v(t)\right))\right)^N,\quad t\in\mathfrak{D},
\end{equation*}
which yields $P(\lambda;\mu,\nu)=(\lambda+\mu)^N$ if $N=1$ or $N=2$.
\end{proof}

As a consequence of this Lemma, if a solution $\lambda(\mu,\nu)$ of the PDE \eqref{PDE} is a zero of a linear or
quadratic polynomial with rational coefficients in $\mu$ and $\nu$, then $\kappa=\frac{1}{2}$ and
$\lambda(\mu,\nu)=-\mu$. In fact, the function $\lambda(\mu,\nu)=-\mu$ solves \eqref{PDE} for $\kappa=\frac{1}{2}$,
but since $\lambda(0,0)=0$ and the spectrum of $A(\frac{1}{2};0,0)$ is given by $\Z\setminus\{0\}$, it is not an
eigenvalue of $A(\frac{1}{2};\mu,\nu)$ for any $(\mu,\nu)\in\R^2$. The following considerations show that this solution
is nevertheless of interest.

\section{Monodromy eigenvalues}

In this section we consider the case that $\kappa$ is a positive half-integer, i.e., $\kappa = k-\frac{1}{2}$ with some
positive integer $k$, and we assume that the matrix $C$ defined in \eqref{Coeff1} has distinct eigenvalues, i.e.,
$\mu^2\neq\nu^2$. For such $\kappa$ and $(\mu,\nu)$ there is in addition to the classical eigenvalues of
$A(\kappa;\mu,\nu)$ another type of ``special values" which we call monodromy eigenvalues. In order to
introduce this concept, we first recall the characterisation of eigenvalues according to Lemma~\ref{Zeros1}: A point
$\lambda$ is an eigenvalue of $A(\kappa;\mu,\nu)$ if and only if the system \eqref{AngSys} has a nontrivial solution of
the form
\begin{equation} \label{Holo}
y(x) = [x(1-x)]^{\frac{\kappa}{2}+\frac{1}{4}}\eta(x)
\end{equation}
where $\eta:\C\longrightarrow\C^2$ is an entire vector function. Now, as the difference of the characteristic values
$\pm\left(\frac{\kappa}{2}+\frac{1}{4}\right)$ at $0$ and $1$ is an integer, the differential equation \eqref{AngSys} has
a fundamental matrix of the form
\begin{equation} \label{Mono}
Y(x) = [x(1-x)]^{-\frac{\kappa}{2}-\frac{1}{4}}H(x)
\end{equation}
where $H(x)=H_a(x)(x-a)^{J_a}$ in $\mathfrak{B}_a$, $a\in\{0,1\}$, with some holomorphic function
$H_a:\mathfrak{B}_a\longrightarrow\MzC$ and a Jordan matrix $J_a$ (see \cite[Theorem 5.6]{Wasow}). Hence, the matrix
function $H$ is in general not holomorphic in $\mathfrak{B}_0\cup\mathfrak{B}_1$ since it involves logarithmic terms. In
the following, a point $\lambda\in\C$ is called \emph{monodromy eigenvalue} of $A(\kappa;\mu,\nu)$ if and only if the
system \eqref{AngSys} has a fundamental matrix of the form \eqref{Mono} with the property that $H:\C\longrightarrow\MzC$
is an entire matrix function. Monodromy eigenvalues are characterised by the following Lemma.

\begin{lemma} \label{WMP}
For a given half-integer $\kappa>0$ and $(\mu,\nu)\in\C^2$ with $\mu^2\neq\nu^2$, a point $\lambda\in\C$ is a
monodromy eigenvalue of $A(\kappa;\mu,\nu)$ if and only if the system \eqref{AngSys} has a nontrivial solution of the
form
\begin{equation} \label{Poly}
[x(1-x)]^{-\frac{\kappa}{2}-\frac{1}{4}}p(x)e^{2tx},
\end{equation}
where $p:\C\longrightarrow\C^2$ is a polynomial vector function and $t = \pm\sqrt{\nu^2-\mu^2}$.
\end{lemma}

\begin{proof}
By means of the transformation $y(x)=x^\alpha(1-x)^\alpha\tilde y(x)$ with $\alpha:=\frac{\kappa}{2}+\frac{1}{4}$,
the differential equation \eqref{AngSys} is equivalent to the system
\begin{equation} \label{Index}
\tilde y'(x) = \left[\frac{1}{x}\,\tilde B_0 + \frac{1}{x-1}\,\tilde B_1 + C\right]\tilde y(x)
\end{equation}
where
\begin{equation} \label{MatB}
\tilde B_0 = \left(\begin{array}{cc} 0 & \mu-\lambda \\[1ex] 0 & k \end{array}\right),\quad
\tilde B_1 = \left(\begin{array}{cc} k & 0 \\[1ex] \mu-\lambda & 0 \end{array}\right).
\end{equation}
Now, if $\lambda$ is a monodromy eigenvalue of $A(\kappa;\mu,\nu)$, then the system \eqref{Index} has a holomorphic
fundamental matrix $H:\C\longrightarrow\MzC$. Since the coefficient matrix of \eqref{Index} is a rational matrix function
which is bounded at infinity, an extension of Halphen's Theorem (see \cite[Theorem 2.4]{GUW}) implies that the system
\eqref{Index} has a fundamental matrix of the form $R(x)e^{Dx}$ with some rational matrix function $R$ and
$D:=\diag(-2\,t,2\,t)$ (note that $\pm 2\,t$ are the eigenvalues of $C$). Moreover, $R(x)e^{Dx}=H(x)Q$ with some
invertible matrix $Q$, and therefore $R(x)=H(x)Qe^{-Dx}$ is an entire matrix function in $\C$. This implies that
$R:\C\longrightarrow\MzC$ is a polynomial. Vice versa, suppose that the system \eqref{AngSys} has a nontrivial solution
$y(x)=[x(1-x)]^{-\alpha}p(x)e^{2tx}$ with some polynomial vector function $p:\C\longrightarrow\C^2$. Defining
\begin{equation*}
\tilde y(x) := e^{-2t}Ky(1-x) = [x(1-x)]^{-\alpha}Kp(1-x)e^{-2tx}
\end{equation*}
with $K$ given by \eqref{K}, it follows that $\tilde y$ is a solution of \eqref{AngSys} which is linearly independent of
$y$. Therefore, \eqref{AngSys} has a fundamental matrix of the type \eqref{Mono}, where
$H(x)=\left(p(x)e^{2tx},Kp(1-x)e^{-2tx}\right)$ is an entire matrix function.
\end{proof}

\begin{theorem} \label{MEV1}
For fixed $\kappa=k-\frac{1}{2}$ with a positive integer $k$ there exists a polynomial $P(\kappa;\lambda;\mu,\nu)$ of
degree $2k-1$ in $\lambda$ with polynomial coefficients in $\mu$ and $\nu$ such that for each $(\mu,\nu)\in\C^2$ with
$\mu^2\neq\nu^2$ a point $\lambda\in\C$ is a monodromy eigenvalue of $A(\kappa;\mu,\nu)$ if and only if $\lambda$ is a
zero of $P(\kappa;\sdot;\mu,\nu)$. Moreover, the integers $1-k,\ldots,k-1$ are the zeros of $P(\kappa;\sdot;0,0)$, and
for $\kappa=\frac{1}{2}$ we obtain $P(\frac{1}{2};\lambda;\mu,\nu)=\lambda+\mu$.
\end{theorem}

\begin{proof}
A point $\lambda$ is a monodromy eigenvalue of $A(\kappa;\mu,\nu)$ if and only if the differential equation \eqref{Index}
has a nontrivial solution $p(x)e^{2tx}$, where $p(x)=\sum_{n=0}^N p_n x^n$, $p_N\neq 0$, is a polynomial vector function,
and $t = \pm\sqrt{\nu^2-\mu^2}$. In the following we assume $t=\sqrt{\nu^2-\mu^2}$ (the main branch of the square root)
but all considerations remain valid if we replace $t$ with $-t$. If we set $\Lambda := \lambda-\mu$ and
$\tilde C := C-t$, then the polynomial $p$ satisfies the differential equation
\begin{equation} \label{PolSys}
p'(x) = \left[\frac{1}{x}\,\tilde B_0 + \frac{1}{x-1}\,\tilde B_1 + \tilde C\right]\,p(x),
\end{equation}
where the coefficient matrices take the form
\begin{equation*}
\tilde B_0 = \left(\begin{array}{cc} 0 & -\Lambda \\[1ex] 0 & k \end{array}\right),\quad
\tilde B_1 = \left(\begin{array}{cc} k & 0 \\[1ex] -\Lambda & 0 \end{array}\right),\quad
\tilde C = \left(\begin{array}{cc} -2\nu-2t & -2\mu \\[1ex] 2\mu & 2\nu-2t \end{array}\right).
\end{equation*}
It is easy to see that the coefficients $p_n\in\C^2$, $n=0,\ldots,N$, form a nontrivial solution of the linear system
of equations
\begin{gather}
\tilde B_0 p_0 = 0,\quad (\tilde B_0 - 1)p_1 + \tilde S p_0 = 0, \label{LinSysA} \\
(\tilde B_0 - n)p_n + (\tilde S + n-1)p_{n-1} - \tilde C p_{n-2} = 0\quad(n=2,\ldots,N), \label{LinSysB} \\
(\tilde S + N)p_N - \tilde C p_{N-1} = 0,\quad - \tilde C p_N = 0, \label{LinSysC}
\end{gather}
where
\begin{equation*}
\tilde S := \tilde C-\tilde B_0-\tilde B_1 = \left(\begin{array}{cc} -2\nu-2t-k & -2\mu+\Lambda \\[1ex] 2\mu+\Lambda & 2\nu-2t-k \end{array}\right).
\end{equation*}
Multiplying the first equation in \eqref{LinSysC} from the left with the matrix $\tilde C+4t$ and observing that
$(\tilde C+4t)\tilde C = 0$, we get
\begin{equation*}
0 = (\tilde C+4t)(\tilde S+N)p_N
  = \left(\begin{array}{cc} N-k & -\Lambda \\[1ex] -\Lambda & N-k \end{array}\right)\tilde C p_N + 4t(N-k)p_N = 4t(N-k)p_N.
\end{equation*}
Since $t\neq 0$ and $p_N\neq 0$, it follows that $N=k$. Due to technical reasons we have to distinguish between the
cases $k\geq 2$ and $k=1$. We will proceed at first with a detailed proof for the more complicated case $k\geq 2$.
Adding the second equation in \eqref{LinSysC} to the first one and then both equations in \eqref{LinSysC} to
\eqref{LinSysB} for $n=N$, we obtain
\begin{gather}
\tilde B_0 p_0 = 0,\quad (\tilde B_0 - 1)p_1 + \tilde S p_0 = 0, \label{LinSys1} \\
(\tilde B_0 - n)p_n + (\tilde S+n-1)p_{n-1} - \tilde C p_{n-2} = 0\quad(n=2,\ldots,k-1), \label{LinSys2} \\
-\tilde B_1 p_k + \left(\begin{array}{cc} -1 & \Lambda \\[1ex] \Lambda & -1 \end{array}\right)p_{k-1} - \tilde C p_{k-2} = 0, \label{LinSys3} \\
\left(\begin{array}{cc} 0 & \Lambda \\[1ex] \Lambda & 0 \end{array}\right) p_k - \tilde C p_{k-1} = 0,\quad - \tilde C p_k = 0. \label{LinSys4}
\end{gather}
The system above consists of $2k+6$ linear equations for $2k+2$ unknowns. In the following we prove that only $2k+2$ of
these equations are linearly independent. Summation of all equations \eqref{LinSysA} -- \eqref{LinSysC} yields
$-\tilde B_1\sum_{n=0}^k p_n = 0$. Because of $\rank(\tilde B_1) = 1$, it is possible to eliminate the second line of
the first equation in \eqref{LinSys3} by means of line transformations, and since also $\rank(\tilde B_0)=1$, we can
delete the first line of the first equation in \eqref{LinSys1}. Thus, the system \eqref{LinSys1} -- \eqref{LinSys4}
consists of at most $2k+4$ linearly independent equations. In order to reduce the equations \eqref{LinSys4} further, we
have to consider the cases $\nu-t\neq 0$ and $\nu-t=0$ separately. First, let us assume that $\nu-t\neq 0$. Multiplying
the equations in \eqref{LinSys4} from the left by the invertible matrix
\begin{equation*}
T := \left(\begin{array}{cc} \nu-t & \mu \\[1ex] 0 & 1 \end{array}\right),
\end{equation*}
it follows that \eqref{LinSys4} is equivalent to
\begin{equation} \label{LinSys4a}
\left(\begin{array}{cc} 0 & 0 \\[1ex] \Lambda & 0 \end{array}\right) p_k - \left(\begin{array}{cc} 0 & 0 \\[1ex] 2\mu & 2\nu-2t \end{array}\right)p_{k-1} = 0,
\quad \left(\begin{array}{cc} 0 & 0 \\[1ex] 2\mu & 2\nu-2t \end{array}\right)p_k = 0.
\end{equation}
Now, we can represent the system of the linear equations \eqref{LinSys1} -- \eqref{LinSys3}, \eqref{LinSys4a} as a matrix
equation $\tilde\Gamma\tilde p = 0$ with $\tilde p = (p_0,\ldots,p_k)\in\C^{2k+2}$ and the
$(2k+2)\times(2k+2)$ matrix
\begin{equation}
\setlength{\arrayrulewidth}{0.1pt}
\setlength{\extrarowheight}{1ex}
\tilde\Gamma := \left(\begin{array}{c|c|c|c|c|c|c}
 \begin{array}{cc} 0 & k \end{array} & \begin{array}{cc} 0 & 0
 \end{array} & \cdots & & & \cdots & \begin{array}{cc} 0 & 0 \end{array} \\[1ex] \hline
 \tilde S & \tilde B_0 - I & 0 & & & & 0 \\
-\tilde C & \tilde S + I & \tilde B_0 - 2\,I & 0 &  & & \vdots \\[1ex]
 0 & -\tilde C & \tilde S + 2\,I & \tilde B_0 - 3\,I & 0 & & \\
 \vdots & \ddots & \ddots & \ddots & \ddots & \ddots & \vdots \\[1ex]
 0 & \cdots & 0 & -\tilde C & \tilde S + (k-2)I & \tilde B_0 - (k-1)I & 0 \\ \hline
 \begin{array}{cc} 0 & 0 \\ 0 & 0 \\ 0 & 0 \end{array} &
 \begin{array}{c} \cdots \\ \cdots \\ \cdots \end{array} &
 \begin{array}{c} \cdots \\ \cdots \\ \cdots \end{array} &
 \begin{array}{cc} 0 & 0 \\ 0 & 0 \\ 0 & 0 \end{array} &
 \begin{array}{cc} 2\nu+2t & 2\mu \\ 0 & 0 \\ 0 & 0 \end{array} &
 \begin{array}{cc} -1 & \Lambda \\ -2\mu & 2t-2\nu \\ 0 & 0 \end{array} &
 \begin{array}{cc} -k & 0 \\ \Lambda & 0 \\ -2\mu & 2t-2\nu \end{array}
\end{array}\right).\label{GammaTilde}
\end{equation}
Let $\hat\Gamma$ be the $(2k+1)\times(2k+1)$-matrix obtained from $\tilde\Gamma$ by deleting the last row and
column. Then $\lambda$ is a monodromy eigenvalue of $A(\kappa;\mu,\nu)$ if and only if
$0 = \det\tilde\Gamma = (2t-2\nu)\det\hat\Gamma$, i.e., $\det\hat\Gamma = 0$ since $\nu-t\neq 0$. Now, suppose that
$\nu-t=0$. We will prove that also in this case $\lambda$ is a monodromy eigenvalue if and only if
$\det\hat\Gamma = 0$. Note that $\nu=t$ implies $\mu = 0$, and therefore the equations in \eqref{LinSys4} are
equivalent to
\begin{equation} \label{LinSys4b}
\left(\begin{array}{cc} 0 & \Lambda \\[1ex] \Lambda & 0 \end{array}\right) p_k + \left(\begin{array}{cc} 4t & 0 \\[1ex] 0 & 0 \end{array}\right)p_{k-1} = 0,
\quad \left(\begin{array}{cc} -4t & 0 \\[1ex] 0 & 0 \end{array}\right)p_k = 0.
\end{equation}
If $\lambda$ is a monodromy eigenvalue of $A(\kappa;\mu,\nu)$, then the vector $\tilde p = (p_0,\ldots,p_k)$ is a
nontrivial solution of the matrix equation $\tilde\Gamma\tilde p = 0$ even though $T$ is not invertible for $\nu-t=0$.
If we assume $\det\hat\Gamma\neq 0$, it follows that $\Lambda\neq 0$ and the first $2k+1$ components of
$\tilde p$ must be zero. In particular, $p_1=\cdots=p_{k-1}=0$, and the first equation in \eqref{LinSys4b} yields
$p_k=0$. Thus $\tilde p=0$, and this contradiction implies $\det\hat\Gamma = 0$. Conversely, if $\det\hat\Gamma = 0$,
then either $\Lambda=0$ and $\tilde p := e_{2k+2}$ (the $2k+2$-nd unit vector in $\mathbb C^{2k+2}$) is a nontrivial
solution of \eqref{LinSys1} -- \eqref{LinSys3} and \eqref{LinSys4b}, or $\Lambda\neq 0$. In the latter case, there exists
a vector $\hat p\neq 0$ with components $\hat p_1,\ldots,\hat p_{2k+1}\in\C$ such that $\hat\Gamma\hat p = 0$. Defining
$q := \frac{4t}{\Lambda}\hat p_{2k-1}$, then $\tilde p := \left(\hat p,q\right)\in\C^{2k+2}$ is a nontrivial solution of
the equations \eqref{LinSys1} -- \eqref{LinSys3} and \eqref{LinSys4b}, i.e., $\lambda$ is a monodromy eigenvalue. Hence,
we have shown that for all $(\mu,\nu)\in\C^2$ with $\mu^2\neq\nu^2$ a point $\lambda\in\C$
is a monodromy eigenvalue of $A(\kappa;\mu,\nu)$ if and only if $\det\hat\Gamma = 0$. In order to prove that
$\det\hat\Gamma$ is a polynomial in $\Lambda$ of degree $2k-1$, we apply once more appropriate line transformations to
$\hat\Gamma$. Adding successively the second to the fourth line, the fourth
to the sixth line and so on up to the $2k$-th line, then $\det\hat\Gamma = \det\Gamma$ with the
$(2k+1)\times(2k+1)$-matrix
\begin{equation*}
\setlength{\arrayrulewidth}{0.1pt}
\setlength{\extrarowheight}{1ex}
\Gamma(\kappa;\Lambda;\mu,\nu;t) := \left(\begin{array}{ccccccc|c}
 \begin{array}{cc} 0 & k \end{array} & \begin{array}{cc} 0 & 0 \end{array} & \hdotsfor{5} &  \multicolumn{1}{|c}{0} \\[1ex] \hline
 \tilde S_0 & \tilde B_0 - I & 0 & & & & & \vdots \\[1ex]
-\tilde R & \tilde S_1 & \tilde B_0 - 2\,I & 0 &  & & &    \\[1ex]
 \tilde Q & -\tilde R & \tilde S_2 & \tilde B_0 - 3\,I & 0 & & &   \\[1ex]
 \tilde Q & \tilde Q & -\tilde R & \tilde S_3 & \tilde B_0 - 4\,I & 0 & & \\
 \vdots & & \ddots & \ddots & \ddots & \ddots & \ddots & \vdots  \\[1ex]
 \tilde Q &  &  & \tilde Q & -\tilde R & \tilde S_{k-2} & \tilde B_0 - (k-1)I & 0 \\[1ex] \hline
 \tilde Q & \hdotsfor{3} & \tilde Q & \tilde Q & -\tilde R & \begin{array}{c} -k \\ \Lambda \end{array}
\end{array}\right)
\end{equation*}
where
\begin{gather*}
\tilde Q := \left(\begin{array}{cc} -k & 0 \\[1ex] 0 & 0 \end{array}\right),\quad
\tilde R := \left(\begin{array}{cc} k & 0 \\[1ex] 2\mu & 2\nu-2t \end{array}\right),\quad
\tilde S_n := \left(\begin{array}{cc} -2\nu-2t-k & -2\mu \\[1ex] 2\mu+\Lambda & 2\nu-2t+n-k \end{array}\right),\quad
n=0,\ldots,k-2.
\end{gather*}
Now, $\Lambda$ appears at most once in each row and each column, whereas only the first and the $2k$-th line
contain no entry involving $\Lambda$. It is easy to verify that $\det\Gamma(\Lambda;\mu,\nu;t)$ has the form
$\pm k^2\Lambda^{2k-1}+\mbox{(terms of lower order in $\Lambda$)}$, and therefore $\det\Gamma(\Lambda;\mu,\nu;t)$ is
a polynomial in $\Lambda$ with polynomial coefficients in $\mu$, $\nu$ and $t$. Moreover, for all
$(\mu,\nu)\in\C^2$ with $\mu^2\neq\nu^2$ a point $\lambda$ is a monodromy eigenvalue of $A(\kappa;\mu,\nu)$ if and only
if the determinant of $\Gamma(\kappa;\Lambda;\mu,\nu;t)$ vanishes. As mentioned at the beginning of the proof, this result
remains valid if we replace $t$ with $-t$. Hence, the zeros of the polynomials $\det\Gamma(\kappa;\Lambda;\mu,\nu;t)$ and
$\det\Gamma(\kappa;\Lambda;\mu,\nu;-t)$
coincide, which implies that $\det\Gamma(\kappa;\Lambda;\mu,\nu;t)=\det\Gamma(\kappa;\Lambda;\mu,\nu;-t)$.
Consequently, the polynomial $P(\kappa;\lambda;\mu,\nu):=\det\Gamma(\kappa;\Lambda;\mu,\nu;t)$ contains no terms in $t$
of odd order, and the terms of even order in $t$ depend only on $t^2=\nu^2-\mu^2$. It follows that $P$ is a polynomial
of degree $2k-1$ in $\lambda$ with polynomial coefficients in $\mu$ and $\nu$, and the zeros of $P$ are exactly the
monodromy eigenvalues of $A(\kappa;\mu,\nu)$.

Next, we prove that the integers $1-k,\ldots,k-1$ are the zeros of the polynomial $P(\kappa;\lambda;0,0)$. To this aim,
let $\Gamma_0$ be the $(2k\times 2k)$ matrix obtained from $\hat\Gamma$ for $(\mu,\nu)=(0,0)$ by deleting the last row and
column. Then
\begin{equation*}
\setlength{\arrayrulewidth}{0.1pt}
\setlength{\extrarowheight}{1ex}
\Gamma_0 = \left(\begin{array}{ccccccc}
 \begin{array}{cc} 0 & k \end{array} & \begin{array}{cc} 0 & 0
 \end{array} & \hdotsfor{4} & 0\quad  0 \\[1ex] \hline
 Q & \tilde B_0 - I & 0 & \hdotsfor{3} & 0 \\[1ex]
 0 & Q + I & \tilde B_0 - 2\,I & 0 & & & \vdots \\[1ex]
 0 & 0 & Q + 2\,I & \tilde B_0 - 3\,I & 0 & & \\
 \vdots & & \ddots & \ddots & \ddots & \ddots & \vdots \\[1ex]
 \vdots & & & 0 & Q+(k-3)I & \tilde B_0-(k-2)I & 0 \\[1ex]
 0 & \hdotsfor{3} & 0 & Q+(k-2)I & \tilde B_0-(k-1)I \\[1ex] \hline
 \begin{array}{cc} 0 & 0 \end{array}  & \hdotsfor{4} &  0\quad  0 &
  -1 \quad \lambda
\end{array}\right),\quad Q := \left(\begin{array}{cc} -k & \lambda \\[1ex] \lambda & -k \end{array}\right),
\end{equation*}
and $\det\hat\Gamma = \lambda\det\Gamma_0$. Moreover, $\det\Gamma_0=0$ if and only if the equation
$\Gamma_0(p_n)_{n=0}^{k-1}=0$ has a nontrivial solution. Such a nontrivial solution is a constant multiple of the
vector given by the recurrence formula
\begin{align*}
p_0 := \left(\begin{array}{cc} (k-1)! \\[1ex] 0 \end{array}\right),\quad p_n & = (n-\tilde B_0)^{-1}(Q+n-1)p_{n-1} \\
& = -\frac{1}{n(k-n)}\left(\begin{array}{cc} (k-n)(k+1-n)-\lambda^2 & \lambda \\[1ex] n\lambda & -n(k+1-n) \end{array}\right)p_{n-1}
\end{align*}
for $n=1,\ldots,k-1$. By induction, it can be shown that
\begin{equation*}
p_n  = (-1)^n\frac{(k-n-1)!}{n!}\prod_{j=1}^{n-1}\left[(k-j)^2-\lambda^2\right]\left(\begin{array}{c} k(k-n)-\lambda^2 \\[1ex] n\lambda \end{array}\right).
\end{equation*}
Multiplying the vector $(p_n)_{n=0}^{k-1}$ from the left by the last line of $\Gamma_0$, we get
\begin{equation*}
0 = (-1)^{k-1}\frac{1}{(k-1)!}\prod_{j=1}^{k-2}\left[(k-j)^2-\lambda^2\right]\left(-k+\lambda^2+(k-1)\lambda^2\right)
  = \frac{k(-1)^k}{(k-1)!}\prod_{j=1}^{k-1}\left[(k-j)^2-\lambda^2\right].
\end{equation*}
Hence, $\det\Gamma_0=0$ if and only if $\lambda^2\in\{1,\ldots,(k-1)^2\}$, and therefore $1-k,\ldots,k-1$ are the zeros
of $P(\kappa;\sdot;0,0)$.

It remains to deal with the case $k=1$, where we have to consider only the equations \eqref{LinSysA} and \eqref{LinSysC}.
Adding both equations in \eqref{LinSysC} to the second equation in \eqref{LinSysA} gives \eqref{LinSys3} with
$p_{-1}:=0$. Hence, we can replace \eqref{LinSysA} -- \eqref{LinSysC} with the linear system of equations consisting of
the first equation in \eqref{LinSys1} and the equations \eqref{LinSys3}, \eqref{LinSys4}. Now, by applying a similar
reduction procedure as in the case $k\geq 2$, we obtain the polynomial
\begin{equation*}
P(\textstyle{\frac{1}{2}};\lambda;\mu,\nu) =
\det\left(\begin{array}{ccc} 0 & 1 & 0 \\ -1 & \Lambda & -1 \\ -2\mu & 2t-2\nu & \Lambda \end{array}\right) = \Lambda+2\mu = \lambda+\mu,
\end{equation*}
whose zero $\lambda=-\mu$ is the uniquely determined monodromy eigenvalue of $A(\frac{1}{2};\mu,\nu)$ for each
$(\mu,\nu)\in\C^2$, $\mu^2\neq\nu^2$.
\end{proof}

\begin{corollary} \label{Sector}
For a fixed half-integer $\kappa=k-\frac{1}{2}$ with a positive integer $k$, there exists a neighbourhood
$\mathfrak{U}\subset\C^2$ of $(0,0)$ such that
$A(\kappa;\mu,\nu)$ has exactly $2k-1$ many monodromy eigenvalues $\lambda_0^j(\kappa;\mu,\nu)$, $j=1-k,\ldots,k-1$, for
all $(\mu,\nu)\in\mathfrak{U}$ with $\mu^2\neq\nu^2$. Moreover, $\lambda_0^j(\kappa;\mu,\nu)$ depends holomorphically on
$(\mu,\nu)$, and $\lim_{(\mu,\nu)\to(0,0)}\lambda_0^j(\kappa;\mu,\nu) = j$. In particular, monodromy eigenvalues
and classical eigenvalues are distinct near $(\mu,\nu)=(0,0)$.
\end{corollary}

\begin{remark} Monodromy eigenvalues also appear in the context of spheroidal wave equations. In \cite[Sec. 3.534]{MS}
they are characterised by a similar property as given in Lemma \ref{WMP}, but they are not specified in detail.
\end{remark}

In view of Theorem \ref{MEV1} and Corollary \ref{Sector} we could alternatively define the monodromy eigenvalues of
$A(\kappa;\mu,\nu)$ to be the zeros of the polynomial $P(\kappa;\sdot;\mu,\nu)$ for each $(\mu,\nu)\in\C^2$ (without the
restriction $\mu^2\neq\nu^2$). Then the monodromy eigenvalues $\lambda_0^j(\kappa;0,0) = j$, $j=1-k,\ldots,k-1$, fill in
the gap of integers appearing in the spectrum of $A(\kappa;0,0)$. Moreover,
$P(\textstyle{\frac{1}{2}};\lambda;\mu,\nu) = \lambda+\mu$ is just the polynomial given by Lemma \ref{AlgSol} in the case
$N=1$, and its zero $\lambda_0^0(\frac{1}{2};\mu,\nu)=-\mu$ satisfies the partial differential equation \eqref{PDE}
for $\kappa=\frac{1}{2}$. In the next section we prove that the monodromy eigenvalues of $A(\kappa;\mu,\nu)$ are
solutions of the PDE \eqref{PDE} for each half-integer $\kappa\in\left\{\frac{1}{2},\frac{3}{2},\frac{5}{2},\ldots\right\}$.

\section{Monodromy Preserving Deformations}

In \cite{JMU1}, \cite{JMU2} and \cite{JM},  Jimbo, Miwa \& Ueno developed a general theory for monodromy
preserving deformations of linear ordinary differential equations with rational coefficients. As a main result, they
proved that the monodromy data (Stokes multipliers, connection matrices and exponents of formal monodromy) do not
depend on the deformation parameters if and only if certain non-linear differential equations, the so-called deformation
equations, are satisfied. This result, however, was proved under the restriction that the characteristic values at
regular singular points do not differ by an integer. On the other hand, in the theory of special functions and in many
physical applications the case where the characteristic values differ by an integer is of great significance. In this section we
consider the isomonodromy problem for linear systems with two fixed regular singular points and coefficients which depend
on one parameter $t$. Assuming that the characteristic values at the singular points are distinct and independent of $t$,
we will show that certain components of the monodromy data are constant with respect to $t$ if a deformation equation of
the type \cite[(1.18)]{JMU1} is satisfied. Since the monodromy components in question determine the existence of solutions
of the form \eqref{Holo} and \eqref{Mono}, they are relevant to monodromy and classical
eigenvalue problems. Applying the results to the system \eqref{AngSys} with an eigenvalue $t$ of $C$ as deformation
parameter, it finally turns out that the deformation equation is in principle the characteristic equation of the partial
differential equation \eqref{PDE}.

We start with some basic facts about parameter-dependent regular singular systems. At first, let us consider a family of
$(2\times 2)$ systems of differential equations
\begin{equation} \label{SysY}
\frac{\partial y}{\partial x}(x,t) = \Phi(x,t)y(x,t),\quad (x,t)\in\left(\mathfrak{B}\setminus\{0\}\right)\times\mathfrak{D},
\end{equation}
in an open disc $\mathfrak{B}\subset\C$ with centre $0$ that depends on a parameter $t$ varying in some real or complex
domain $\mathfrak{D}$. It is assumed that \eqref{SysY} has a regular singular point at $0$ for all $t\in\mathfrak{D}$.
More precisely, we suppose that the coefficient matrix $\Phi$ of \eqref{SysY} has the following
properties:
\begin{clist}{(x)}\renewcommand{\thenum}{(\alph{num})}
\item $\Phi(x,t)=\frac{1}{x}\Psi(x,t)$, where $\Psi:\mathfrak{B}\times\mathfrak{D}\longrightarrow\MzC$ is an analytical
matrix function.
\item The eigenvalues $\alpha$ and $\beta$ of $\Psi(0,t)$ are distinct and independent of $t\in\mathfrak{D}$; moreover,
$\re\alpha\leq\re\beta$.
\item There is an analytical function $G:\mathfrak{D}\longrightarrow\MzC$ such that $G(t)$ is invertible and
\begin{equation*}
G(t)^{-1}\Psi(0,t)G(t) = \diag(\alpha,\beta) =: D,\quad t\in\mathfrak{D},
\end{equation*}
\end{clist}
Note that such a matrix function $G$ always exists since the eigenvalues of
$\Psi(0,t)$ are distinct (see \cite[Chap. VII, Sec. 25, Theorem 25.1]{Wasow}).

\begin{lemma} \label{Fundamental}
If the conditions \Hyp{a} -- \Hyp{c} are satisfied, then the system \eqref{SysY} has a fundamental matrix of the form
\begin{equation*}
Y(x,t) = G(t) H(x,t) x^D x^{J(t)}
\end{equation*}
where $H:\mathfrak{B}\times\mathfrak{D}\longrightarrow\MzC$ is analytic, $H(0,t)=I$ for all $t\in\mathfrak{D}$, and
\begin{equation} \label{J}
J(t) = \left(\begin{array}{cc} 0 & 0 \\[1ex] p(t) & 0 \end{array}\right)
\end{equation}
with some analytical function $p:\mathfrak{D}\longrightarrow\C$. Moreover, if $\beta-\alpha$ is not an integer, then
$p\equiv 0$.
\end{lemma}

\begin{proof}
If $\beta-\alpha$ is not an integer, the existence of such a fundamental matrix with $J\equiv 0$ is well known cf.
\cite{SchaefkeFW}. Hence, we have to consider only the case that $k:=\beta-\alpha$ is a positive integer.
By the transformation
\begin{equation} \label{Reduc}
y(x,t) = x^\alpha G(t)y_0(x,t),
\end{equation}
the system \eqref{SysY} is equivalent to the differential equation
\begin{equation} \label{Sys0}
x\,\frac{\partial y_0}{\partial x}(x,t) = \Psi_0(x,t)y_0(x,t),\quad (x,t)\in\left(\mathfrak{B}\setminus\{0\}\right)\times\mathfrak{D},
\end{equation}
where $\Psi_0(x,t) := G(t)^{-1}\Psi(x,t)G(t)-\alpha$ is an analytical matrix function,
\begin{equation*}
\Psi_0(x,t) = \sum_{n=0}^\infty x^n\Psi_{0,n}(t),\quad (x,t)\in\mathfrak{B}\times\mathfrak{D},
\end{equation*}
with $\Psi_{0,0}(t)=\diag(0,k)$ for all $t\in\mathfrak{D}$. Now, for $j=1,\ldots,k-1$ we recursively apply the
transformations
\begin{equation} \label{Trafo}
y_{j-1}(x,t) = \left(\begin{array}{cc} 1 & 0 \\[1ex] \frac{x}{j-k}\psi_{j-1}(t) & x \end{array}\right)y_j(x,t),
\end{equation}
where $\psi_{j-1}$ denotes the $(2,1)$-coefficient of the matrix $\Psi_{j-1,1}$. At each step, $y_j(x,t)$ is a solution
of a system
\begin{equation} \label{Sys1}
x\,\frac{\partial y_j}{\partial x}(x,t) = \Psi_j(x,t)y_j(x,t),\quad (x,t)\in\left(\mathfrak{B}\setminus\{0\}\right)\times\mathfrak{D},
\end{equation}
where the coefficient matrix $\Psi_j$ is analytic in $\mathfrak{B}\times\mathfrak{D}$,
\begin{equation*}
\Psi_j(x,t) = \sum_{n=0}^\infty x^n \Psi_{j,n}(t),\quad (x,t)\in\mathfrak{B}\times\mathfrak{D},
\end{equation*}
with $\Psi_{j,0}(t)=\diag(0,k-j)$ for all $t\in\mathfrak{D}$, and $\Psi_{j,n}(t)$, $n=1,\ldots,j-1$, are
lower triangular matrix functions (that means, the $(1,2)$ component is identically zero). Finally, by the shearing
transformation
\begin{equation} \label{Shear}
y_{k-1}(x,t) = \left(\begin{array}{cc} 1 & 0 \\[1ex] 0 & x \end{array}\right)y_k(x,t),
\end{equation}
we obtain a differential equation
\begin{equation} \label{Sys2}
x\,\frac{\partial y_k}{\partial x}(x,t) = \Psi_k(x,t)y_k(x,t),\quad (x,t)\in\left(\mathfrak{B}\setminus\{0\}\right)\times\mathfrak{D},
\end{equation}
where $\Psi_k:\mathfrak{B}\times\mathfrak{D}\longrightarrow\MzC$ is an analytical matrix function,
\begin{equation*}
\Psi_k(x,t) = \sum_{n=0}^\infty x^n \Psi_{k,n}(t),\quad (x,t)\in\mathfrak{B}\times\mathfrak{D},
\end{equation*}
satisfying
\begin{equation*}
\Psi_{k,0}(t) = \left(\begin{array}{cc} 0 & 0 \\[1ex] p(t) & 0 \end{array}\right) =: J(t),\quad t\in\mathfrak{D},
\end{equation*}
with some analytical function $p:\mathfrak{D}\longrightarrow\C$. Note that $p$ is just the $(2,1)$-component of
$\Psi_{k-1,1}$. Moreover, $\Psi_{k,n}(t)$, $n=0,\ldots,k$, are lower triangular matrices for all $t\in\mathfrak{D}$.
Now, the system \eqref{Sys2} has a fundamental matrix of the form
\begin{equation*}
\tilde Y(x,t) = \tilde H(x,t) x^{J(t)}
\end{equation*}
provided that $\tilde H$ is a solution of the matrix differential equation
\begin{equation} \label{SysH}
x\,\frac{\partial\tilde H}{\partial x}(x,t) = \Psi_k(x,t)\tilde H(x,t)-\tilde H(x,t)J(t),\quad (x,t)\in\mathfrak{B}\times\mathfrak{D},
\end{equation}
such that for each $t\in\mathfrak{D}$ the matrix $\tilde H(x,t)$ is invertible for some, and hence all,
$x\in\mathfrak{B}$. Obviously, \eqref{SysH} has a formal solution
\begin{equation} \label{SolH}
\tilde H(x,t) = \sum_{n=0}^\infty x^n \tilde H_n(t),\quad (x,t)\in\mathfrak{B}\times\mathfrak{D},
\end{equation}
where $\tilde H_0(t)=I$ and the coefficients $\tilde H_n$, $n>0$, are uniquely determined by the recurrence relation
\begin{equation}
\left(J(t)-n\right)\tilde H_n(t) - \tilde H_n(t)J(t) = -\sum_{j=0}^{n-1}\Psi_{k,n-j}(t)\tilde H_j(t)
\end{equation}
Following the proof of Theorem 5.3 in the book of Wasow \cite{Wasow}, it can be shown that the series \eqref{SolH}
converges uniformly in every compact subset of $\mathfrak{B}\times\mathfrak{D}$. Thus, a Weierstrass theorem implies
that $\tilde H$ is analytic in $\mathfrak{B}\times\mathfrak{D}$, and therefore $\tilde H$ is an actual solution of
\eqref{SysH}. Further, since $J(t)$ has the special form \eqref{J} and $\Psi_{k,j}(t)$, $j=0,\ldots,k$, are lower
triangular matrices, it is easy to
verify that $\tilde H_j(t)$ are lower triangular matrices for $j = 0,\ldots,k$. Now, by combining the transformations
\eqref{Reduc}, \eqref{Trafo} and \eqref{Shear}, it follows that the differential equation \eqref{SysY} has a fundamental
matrix of the form
\begin{equation} \label{FM}
Y(x,t) = x^\alpha G(t)\left(\begin{array}{cc} 1 & 0 \\[1ex] x\,q(x,t) & x^k \end{array}\right)\tilde H(x,t) x^{J(t)},
\end{equation}
where $q(x,t)$ is a polynomial in $x$ of degree $n-1$ with coefficients depending analytically on $t$,
and $\tilde H(x,t)$ is an analytical matrix function of the type
\begin{equation*}
\tilde H(x,t) = \left(\begin{array}{cc} h_{11}(x,t) & x^{k+1} h_{12}(x,t) \\[1ex] h_{21}(x,t) & h_{22}(x,t) \end{array}\right)
\end{equation*}
satisfying $h_{11}(0,t)=h_{22}(0,t)=1$. Now, if we define
\begin{equation*}
H(x,t) := \left(\begin{array}{cc} h_{11}(x,t) & x\,h_{12}(x,t) \\[1ex] x\,q(x,t)h_{11}(x,t) + x^{k+1}h_{21}(t) & x^2 q(x,t)h_{12}(x,t)+h_{22}(x,t) \end{array}\right),
\end{equation*}
then $H:\mathfrak{B}\times\mathfrak{D}\longrightarrow\MzC$ is analytic, $H(0,t)=I$ for all $t\in\mathfrak{D}$, and
\begin{equation*}
\left(\begin{array}{cc} 1 & 0 \\[1ex] x q(x,t) & x^k \end{array}\right)\tilde H(x,t)
= H(x,t)\left(\begin{array}{cc} 1 & 0 \\[1ex] 0 & x^k \end{array}\right).
\end{equation*}
Hence, we can write the fundamental matrix \eqref{FM} in the form $Y(x,t) = G(t) H(x,t) x^D x^{J(t)}$, where $H$ has
the properties stated in the Lemma.
\end{proof}

Now, we consider a family of $(2\times 2)$ differential systems
\begin{equation} \label{Sys01}
\frac{\partial y}{\partial x}(x,t) = \Phi(x,t)y(x,t),\quad (x,t)\in\left(\mathfrak{G}\setminus\{0,1\}\right)\times\mathfrak{D},
\end{equation}
in a domain $\mathfrak{G}$, $\mathfrak{B}_0\cup\mathfrak{B}_1\subset\mathfrak{G}\subset\C$, with regular singular points
at $x=0$ and $x=1$ and a parameter $t$ varying in some domain $\mathfrak{D}\subset\R$ or $\mathfrak{D}\subset\C$.
Further, we assume that the coefficient matrix $\Phi$ in \eqref{Sys01} has the form
\begin{equation*}
\Phi(x,t)=\frac{1}{x(x-1)}\Psi(x,t),\quad (x,t)\in\left(\mathfrak{G}\setminus\{0,1\}\right)\times\mathfrak{D},
\end{equation*}
where $\Psi:\mathfrak{G}\times\mathfrak{D}\longrightarrow\MzC$ is an analytical matrix function with the following
properties:
\begin{clist}{(II)}\renewcommand{\thenum}{(\Roman{num})}
\item The eigenvalues $\alpha$, $\beta$ of $\Psi(a,t)$ are distinct and independent of $t\in\mathfrak{D}$
and $a\in\{0,1\}$; in addition, $\re\alpha\leq\re\beta$.
\item There are analytical functions $G_a:\mathfrak{D}\longrightarrow\MzC$, $a\in\{0,1\}$, such that $G_a(t)$ is
invertible for all $t\in\mathfrak{D}$ and
\begin{equation*}
G_a(t)^{-1}\Psi(a,t)G_a(t) = (-1)^a \diag(\alpha,\beta).
\end{equation*}
\end{clist}
From Lemma \ref{Fundamental} it follows that the system \eqref{Sys01} possesses a fundamental matrix of the form
\begin{equation} \label{FM01}
Y_a(x,t) = G_a(t) H_a(x,t)(x-a)^D(x-a)^{J_a(t)}
\end{equation}
in the unit disc $\mathfrak{B}_a\subset\mathfrak{G}$ with centre $a\in\{0,1\}$, where
$H_a:\mathfrak{B}_a\times\mathfrak{D}\longrightarrow\MzC$ is an analytical matrix function satisfying
$H_a(0,t)=I$ for all $t\in\mathfrak{D}$, $D = \diag(\alpha,\beta)$, and
\begin{equation} \label{Jordan}
J_a(t) = \left(\begin{array}{cc} 0 & 0 \\[1ex] p_a(t) & 0 \end{array}\right)
\end{equation}
with some analytical function $p_a:\mathfrak{D}\longrightarrow\C$. By analytic continuation along curves, we can
assume that $Y_a$ is defined on the universal covering $\mathfrak{R}$ of the set $\mathfrak{G}\setminus\{0,1\}$.
Since $Y_a(x\,e^{2i\pi}+a,t)=Y_a(x+a,t)e^{2i\pi D}\left[I+2\pi i J_a(t)\right]$ for all
$(x,t)\in\left(\mathfrak{B}_0\setminus\{0\}\right)\times\mathfrak{D}$, the diagonal matrix $D$ and the Jordan type matrix
$J_a(t)$ represent the monodromy behaviour of $Y_a$ corresponding to a circuit around $a\in\{0,1\}$. Moreover, as $Y_0$
and $Y_1$ are both fundamental matrices of the same differential equation \eqref{Sys01}, there
exists an analytical matrix function $Q:\mathfrak{D}\longrightarrow\MzC$ such that $Y_0(x,t)=Y_1(x,t)Q(t)$ for all
$(x,t)\in\left(\mathfrak{G}\setminus\{0,1\}\right)\times\mathfrak{D}$, which is called the connection matrix for $Y_0$
and $Y_1$. The next result gives a sufficient condition that certain components of the monodromy data $J_a$ and $Q$ are
constant in $\mathfrak{D}$. For this reason, we establish in addition to \Hyp{I} -- \Hyp{II} the following assumptions on
the coefficient matrix $\Phi$:
\begin{clist}{(III)}\setcounter{num}{2}\renewcommand{\thenum}{(\Roman{num})}
\item There exists an analytical function $\Omega:\mathfrak{G}\times\mathfrak{D}\longrightarrow\MzC$ such that
\begin{equation} \label{DefEqu}
\frac{\partial\Phi}{\partial t}(x,t) + \Phi(x,t)\Omega(x,t) = \Omega(x,t)\Phi(x,t) + \frac{\partial\Omega}{\partial x}(x,t),
\quad (x,t)\in(\mathfrak{G}\setminus\{0,1\})\times\mathfrak{D},
\end{equation}
\item The matrix functions $G_a$, $a\in\{0,1\}$, satisfy the linear differential equations
\begin{equation} \label{GauEqu}
\frac{\partial G_a}{\partial t}(t) = \Omega(a,t)G_a(t),\quad t\in\mathfrak{D}.
\end{equation}
\end{clist}

\begin{theorem} \label{Monodromy}
If the conditions \Hyp{I} -- \Hyp{IV} are satisfied, then
\begin{equation}
\frac{\partial J_a}{\partial t}\equiv\frac{\partial Q_{21}}{\partial t}\equiv 0\mbox{ in }\mathfrak{D},
\end{equation}
where the Jordan matrices $J_a$, $a\in\{0,1\}$, are given by \eqref{Jordan} and $Q_{12}:\mathfrak{D}\longrightarrow\C$
denotes the $(1,2)$-component of the connection matrix $Q$ for $Y_0$ and $Y_1$.
\end{theorem}

\begin{proof}
Let $\gamma := \beta-\alpha$, and for fixed $a\in\{0,1\}$ we define
\begin{equation*}
Z_a(x,t) := \frac{\partial Y_a}{\partial t}(x,t)-\Omega(x,t)Y_a(x,t),\quad (x,t)\in\mathfrak{R}\times\mathfrak{D}.
\end{equation*}
From \eqref{Sys01} and the deformation equation \eqref{DefEqu} it follows that
\begin{align*}
\frac{\partial Z_a}{\partial x}
& = \frac{\partial^2 Y_a}{\partial x\,\partial t} - \frac{\partial\Omega}{\partial x}\,Y_a - \Omega\,\frac{\partial Y_a}{\partial x}
  = \frac{\partial (\Phi\,Y_a)}{\partial t} - \frac{\partial\Omega}{\partial x}\,Y_a - \Omega\,\Phi\,Y_a \\
& = \Phi\,\frac{\partial Y_a}{\partial t} + \left(\frac{\partial\Phi}{\partial t} - \frac{\partial\Omega}{\partial x} - \Omega\,\Phi\right)Y_a
  = \Phi\left(\frac{\partial Y_a}{\partial t}-\Omega\,Y_a\right) = \Phi\,Z_a,
\end{align*}
and therefore $Z_a$ is a matrix solution of the differential equation \eqref{Sys01} in $\mathfrak{R}$. Hence,
there exists an analytical function $C_a:\mathfrak{D}\longrightarrow\MzC$ such that
\begin{equation*}
Z_a(x,t) = Y_a(x,t)C_a(t),\quad (x,t)\in\mathfrak{R}\times\mathfrak{D}.
\end{equation*}
Now, by means of the differential equation \eqref{GauEqu}, we get
\begin{align*}
\frac{\partial Y_a}{\partial t}
& = \frac{\partial G_a}{\partial t}H_a(x-a)^D(x-a)^{J_a} + G_a\frac{\partial H_a}{\partial t}(x-a)^D(x-a)^{J_a}
    + \log(x-a)\,G_a H_a (x-a)^D \frac{\partial J_a}{\partial t} (x-a)^{J_a} \\
& = \left(\Omega(a,\sdot)\,G_a H_a + G_a\frac{\partial H_a}{\partial t} + (x-a)^\gamma\log(x-a)\,G_a H_a\frac{\partial J_a}{\partial t}\right)(x-a)^D(x-a)^{J_a},
\end{align*}
and since $C_a = Y_a^{-1}Z_a$, it results that
\begin{align} \label{C1}
(x-a)^D(x-a)^{J_a} & C_a(x-a)^{-J_a}(x-a)^{-D}  \notag \\
& = (x-a)^D(x-a)^{J_a}Y_a^{-1}\left(\frac{\partial Y_a}{\partial t}-\Omega\,Y_a\right)(x-a)^{-J_a}(x-a)^{-D} \notag \\
& = H_a^{-1}G_a^{-1}(\Omega(a,\sdot)-\Omega)\,G_a H_a + H_a^{-1}\frac{\partial H_a}{\partial t} + (x-a)^\gamma\log(x-a)\,\frac{\partial J_a}{\partial t} \notag \\
& = (x-a)F_a + (x-a)^\gamma\log(x-a)\left(\begin{array}{cc} 0 & 0 \\[1ex] \frac{\partial p_a}{\partial t} & 0 \end{array}\right)
\end{align}
with some analytical function $F_a:\mathfrak{G}\times\mathfrak{D}\longrightarrow\MzC$. Further, by setting
\begin{equation*}
C_a(t) =: \left(\begin{array}{cc} C_{11}(t) & C_{12}(t) \\[1ex] C_{21}(t) & C_{22}(t) \end{array}\right),\quad t\in\mathfrak{D},
\end{equation*}
(for clarity, we omit the index $a$ in the entries of $C_a$), we have
\begin{align} \label{C2}
& (x-a)^D(x-a)^{J_a}C_a(x-a)^{-J_a}(x-a)^{-D} \notag \\
& \qquad = \left(\begin{array}{cc} 1 & 0 \\[1ex] (x-a)^\gamma\log(x-a)\,p_a & (x-a)^\gamma \end{array}\right)
    \left(\begin{array}{cc} C_{11} & C_{12} \\[1ex] C_{21} & C_{22} \end{array}\right)
    \left(\begin{array}{cc} 1 & 0 \\[1ex] -\log(x-a)\,p_a & (x-a)^{-\gamma} \end{array}\right) \notag \\
& \qquad = \left(\begin{array}{cc} C_{11}-\log(x-a)\,p_a C_{12} & (x-a)^{-\gamma}C_{12} \\[1ex] \ast & C_{22}+\log(x-a)\,p_a C_{12} \end{array}\right).
\end{align}
Comparing \eqref{C1} to \eqref{C2}, it follows that $C_{12}\equiv 0$ in $\mathfrak{D}$ since the function in \eqref{C1}
is bounded at $x=a$. This in turn implies $C_{11}\equiv C_{22}\equiv 0$ as the diagonal entries in \eqref{C1} have
a zero at $x=a$ for all $t\in\mathfrak{D}$. Finally, we obtain that
\begin{equation} \label{C3}
(x-a)^D(x-a)^{J_a}C_a(x-a)^{-J_a}(x-a)^{-D} = \left(\begin{array}{cc} 0 & 0 \\[1ex] (x-a)^\gamma C_{21} & 0 \end{array}\right)
\end{equation}
has no logarithmic singularity at $x=a$ and therefore the last term \eqref{C1} vanishes identically. Hence,
$\frac{\partial J_a}{\partial t}\equiv 0$ in $\mathfrak{D}$.

Next, we prove that $\frac{\partial Q_{12}}{\partial t}\equiv 0$. Since $Y_0(x,t)=Y_1(x,t)Q(t)$, it follows that
\begin{equation} \label{Q1}
\frac{\partial Y_0}{\partial t} = \frac{\partial Y_1}{\partial t}\,Q + Y_1\,\frac{\partial Q}{\partial t}
\end{equation}
Further, from $Z_a(x,t)=Y_a(x,t)C_a(t)$ we get
\begin{equation} \label{Q2}
\frac{\partial Y_0}{\partial t}\,Y_0^{-1} - \Omega = Y_0\,C_0\,Y_0^{-1},\quad
\frac{\partial Y_1}{\partial t}\,Y_1^{-1} - \Omega = Y_1\,C_1\,Y_1^{-1}.
\end{equation}
By means of \eqref{Q1} and $Y_0^{-1} = Q^{-1}Y_1^{-1}$, the first equation in \eqref{Q2} becomes
\begin{equation} \label{Q3}
\frac{\partial Y_1}{\partial t}\,Y_1^{-1} - \Omega = Y_1\,Q\,C_0\,Q^{-1}Y_1^{-1} - Y_1\,\frac{\partial Q}{\partial t}\,Q^{-1}Y_1^{-1}.
\end{equation}
Now, \eqref{Q3} and the second equation in \eqref{Q2} imply
\begin{equation*}
Y_1\,C_1\,Y_1^{-1} = Y_1\,Q\,C_0\,Q^{-1}Y_1^{-1} - Y_1\,\frac{\partial Q}{\partial t}\,Q^{-1}Y_1^{-1}
\end{equation*}
and therefore
\begin{equation} \label{Connect}
\frac{\partial Q}{\partial t} = Q\,C_0 - C_1\,Q.
\end{equation}
Note that the matrix function $C_a$ has the form
\begin{equation*}
C_a(t) = \left(\begin{array}{cc} 0 & 0 \\[1ex] c_a(t) & 0 \end{array}\right),\quad a\in\{0,1\}.
\end{equation*}
Hence, if we set
\begin{equation*}
Q(t) =: \left(\begin{array}{cc} Q_{11}(t) & Q_{12}(t) \\[1ex] Q_{21}(t) & Q_{22}(t) \end{array}\right),\quad t\in\mathfrak{D},
\end{equation*}
then \eqref{Connect} is equivalent to the system
\begin{equation*}
\frac{\partial}{\partial t}\left(\begin{array}{cc} Q_{11} & Q_{12} \\[1ex] Q_{21} & Q_{22} \end{array}\right)
= \left(\begin{array}{cc} c_0 Q_{12} & 0 \\[1ex] c_0 Q_{22} - c_1 Q_{11} & c_1 Q_{12} \end{array}\right),
\end{equation*}
and we immediately obtain that $\frac{\partial Q_{12}}{\partial t}\equiv 0$ in $\mathfrak{D}$.
\end{proof}

In the following we apply the results of Lemma \ref{Fundamental} and Theorem \ref{Monodromy} to a family of $(2\times 2)$
differential systems
\begin{equation} \label{ABC}
\frac{\partial y}{\partial x}(x,t) = \left[\frac{1}{x}\,B_0(t) + \frac{1}{x-1}\,B_1(t) + C(t)\right]y(x,t),\quad (x,t)\in\left(\C\setminus\{0,1\}\right)\times\mathfrak{D},
\end{equation}
where $t\in\mathfrak{D}$ with some domain $\mathfrak{D}\subset\R$ or $\mathfrak{D}\subset\C$, and the coefficients
$B_0,\,B_1,\,C:\mathfrak{D}\longrightarrow\MzC$ are supposed to be analytical functions. Further, we assume that the
following conditions hold:
\begin{clist}{(iii)}\renewcommand{\thenum}{(\roman{num})}
\item The eigenvalues $\alpha$, $\beta$ of $B_0(t)$ are distinct and independent of $t\in\mathfrak{D}$. Moreover, they
coincide with the eigenvalues of $B_1(t)$, and $\re\alpha\leq\re\beta$.
\item There are analytical functions $G_a:\mathfrak{D}\longrightarrow\MzC$, $a\in\{0,1\}$, such that $G_a(t)$ is
invertible and
\begin{equation*}
G_0(t)^{-1}B_0(t)G_0(t) = -G_1(t)^{-1}B_1(t)G_1(t) = \diag(\alpha,\beta) =: D,\quad t\in\mathfrak{D}.
\end{equation*}
\end{clist}
Let $Y_a$, $a\in\{0,1\}$, be fundamental matrices of \eqref{ABC} in the open disc $\mathfrak{B}_a\subset\C$ with centre
$a$ and radius $1$ having the form \eqref{FM01}, where $H_a:\mathfrak{B}\times\mathfrak{D}\longrightarrow\MzC$ is
analytical, $H_a(0,t)=I$ for all $t\in\mathfrak{D}$, and $J_a(t)$ is given by \eqref{Jordan} with some analytical
function $p_a:\mathfrak{D}\longrightarrow\C$. Again, by analytic continuation, we assume that $Y_a$ is defined on the
universal covering $\mathfrak{R}$ of $\C\setminus\{0,1\}$, and we denote by $Q:\mathfrak{D}\longrightarrow\MzC$ the
connection matrix for $Y_0$ and $Y_1$. In the sequel we are looking for conditions such that for fixed $t\in\mathfrak{D}$
the system \eqref{ABC} has one of the following properties:
\begin{plist}
\item[\textbf{(P)}] There exists a fundamental matrix $Y$ of the form
\begin{equation} \label{P}
Y(x) = [x(1-x)]^\alpha P(x)e^{C(t)x},\quad x\in\C\setminus\{0,1\},
\end{equation}
where $P:\C\longrightarrow\MzC$ is a polynomial matrix function.
\item[\textbf{(H)}] There exists a nontrivial solution $y$ of the form
\begin{equation} \label{H}
y(x) = [x(1-x)]^\beta h(x),\quad x\in\C\setminus\{0,1\},
\end{equation}
where $h:\C\longrightarrow\C^2$ is an entire vector function.
\end{plist}

\begin{lemma} \label{Property}
Suppose that the conditions \Hyp{i} and \Hyp{ii} are satisfied, and let $t\in\mathfrak{D}$. Then the system \eqref{ABC}
has the property \Hyp{P} if and only if $\beta-\alpha$ is an integer and $p_0(t)=p_1(t)=0$, and it has the property
\Hyp{H} if and only if $Q_{21}(t)=0$.
\end{lemma}

\begin{proof}
By means of the transformation $y(x)=x^\alpha(x-1)^\alpha\tilde y(x)$, the differential equation
\eqref{ABC} is equivalent to the system
\begin{equation} \label{Trans}
\frac{\partial\tilde y}{\partial x}(x,t) = \left[\frac{1}{x}\,\tilde B_0(t) + \frac{1}{x-1}\,\tilde B_1(t) + C(t)\right]\tilde y(x,t),\quad (x,t)\in\left(\C\setminus\{0,1\}\right)\times\mathfrak{D},
\end{equation}
where $\tilde B_0(t) := B_0(t)-\alpha$ and $\tilde B_1(t) := B_1(t)-\alpha$. Moreover,
\begin{equation} \label{FMa}
\tilde Y_a(x,t) = G_a(t) H_a(x,t)\left(\begin{array}{cc} 1 & 0 \\[1ex] 0 & (x-a)^{\beta-\alpha} \end{array}\right)(x-a)^{J_a(t)}
\end{equation}
are fundamental matrices of \eqref{Trans} in a neighbourhood of $a\in\{0,1\}$. First, suppose that $\beta-\alpha$ is an
integer and that $p_0(t)=p_1(t)=0$ holds. In this case $J_0(t)=J_1(t)=0$, and the system \eqref{Trans} has a holomorphic
fundamental matrix since $(x-a)^{\beta-\alpha}$ is holomorphic and $\tilde Y_0(\sdot,t)$, $\tilde Y_1(\sdot,t)$ contain
no logarithmic terms. Moreover, as the coefficient matrix $\Phi(\sdot,t)$ of \eqref{ABC} is a rational function which is
bounded at infinity, the extension of Halphen's Theorem \cite[Theorem 2.4]{GUW} implies that the system \eqref{Trans} has
a fundamental matrix of the form $\tilde Y(x) = R(x)e^{C(t)x}$ with some rational (and hence polynomial) matrix function
$R:\C\longrightarrow\MzC$. Conversely, if \eqref{ABC} has a fundamental matrix of the form \eqref{P}, then
$\tilde Y_0(\sdot,t)$ and $\tilde Y_1(\sdot,t)$ are holomorphic matrix functions, which gives $\beta-\alpha\in\Z$ and
$J_0(t)=J_1(t)=0$.

Next, let us assume that $Q_{12}(t)=0$. If we define
\begin{equation*}
y(x) := Y_0(x,t)\left(\begin{array}{c} 0 \\[1ex] 1 \end{array}\right) = x^\beta G_0(t)H_0(x,t)\left(\begin{array}{c} 0 \\[1ex] 1 \end{array}\right),
\end{equation*}
then $y$ is a nontrivial solution of \eqref{ABC}, and $x^{-\beta}y(x)$ is analytic at $x=0$. Since
$Y_0(x,t)=Y_1(x,t)Q(t)$ and $Q_{12}(t)=0$, we obtain
\begin{equation*}
y(x) = Y_1(x,t)Q(t)\left(\begin{array}{c} 0 \\[1ex] 1 \end{array}\right) = (x-1)^\beta G_1(t)H_1(x,t)\left(\begin{array}{c} 0 \\[1ex] Q_{22}(t) \end{array}\right),
\end{equation*}
and therefore $(1-x)^{-\beta}y(x)$ is analytic in a neighbourhood of $x=1$. Now, by the existence and uniqueness
theorem, $h(x) := [x(1-x)]^{-\beta}y(x)$ can be extended to an entire vector function. Conversely, suppose that
\eqref{ABC} has a nontrivial solution of the form \eqref{H}. Then
\begin{equation*}
y(x) = Y_0(x,t)\left(\begin{array}{c} 0 \\[1ex] c_0 \end{array}\right) = Y_1(x,t)\left(\begin{array}{c} 0 \\[1ex] c_1 \end{array}\right)
\end{equation*}
with some constants $c_0,\,c_1\in\C\setminus\{0\}$. Since $Y_0(x,t)=Y_1(x,t)Q(t)$, it follows that
\begin{equation*}
Q(t)\left(\begin{array}{c} 0 \\[1ex] c_0 \end{array}\right) = \left(\begin{array}{c} 0 \\[1ex] c_1 \end{array}\right),
\end{equation*}
which gives $Q_{21}(t)=0$.
\end{proof}

Now, in addition to \Hyp{i} -- \Hyp{ii}, we assume that the coefficients of \eqref{ABC} satisfy the following
conditions:
\begin{clist}{(iii)}\setcounter{num}{2}\renewcommand{\thenum}{(\roman{num})}
\item There exists an analytical function $\Omega:\C\times\mathfrak{D}\longrightarrow\MzC$ such that the
deformation equation \eqref{DefEqu} holds in $(\C\setminus\{0,1\})\times\mathfrak{D}$, where $\Phi$ is given by
\begin{equation*}
\Phi(x,t) := \frac{1}{x}\,B_0(t) + \frac{1}{x-1}\,B_1(t) + C(t),\quad (x,t)\in\left(\C\setminus\{0,1\}\right)\times\mathfrak{D}.
\end{equation*}
\item The matrix functions $G_a$, $a\in\{0,1\}$, satisfy the differential equations
\begin{equation*}
\frac{\partial G_a}{\partial t}(t) = \Omega(a,t)G_a(t),\quad t\in\mathfrak{D}.
\end{equation*}
\end{clist}
The next result is an immediate consequence of Theorem \ref{Monodromy} and Lemma \ref{Property}.

\begin{corollary} \label{Isomono}
Suppose that the conditions \Hyp{i} -- \Hyp{iv} are satisfied. If \Hyp{P} holds for one $t_0\in\mathfrak{D}$, then
\eqref{ABC} has the property \Hyp{P} for all $t\in\mathfrak{D}$, and if \Hyp{H} holds for one $t_0\in\mathfrak{D}$,
then \eqref{ABC} has the property \Hyp{H} for all $t\in\mathfrak{D}$.
\end{corollary}

Finally, we apply the results of this section to prove that the classical as well as the monodromy eigenvalues of
the Chandrasekhar-Page angular equation in dependence of $(\mu,\nu)\in\R^2$ are (locally) solutions of the partial
differential equation \eqref{PDE}.

\begin{lemma} \label{MPD}
Let $\kappa\in[\frac{1}{2},\infty)$ and $\sigma\in\{-1,+1\}$ be fixed. Moreover, suppose that the functions
$v:\mathfrak{D}\longrightarrow\R\setminus\{0\}$ and $w:\mathfrak{D}\longrightarrow\R$ are solutions of the system
\eqref{Char1} -- \eqref{Char2} on some interval $\mathfrak{D}\subset(0,\infty)$. Finally, let
\begin{equation} \label{Coords}
\mu(t) := \frac{t}{2}\left(v(t)+\frac{\sigma}{v(t)}\right),\quad\nu(t) := \frac{t}{2}\left(v(t)-\frac{\sigma}{v(t)}\right),\quad t\in\mathfrak{D},
\end{equation}
and $t_0\in\mathfrak{D}$. If $w(t_0)$ is an eigenvalue of $A\left(\kappa;\mu(t_0),\nu(t_0)\right)$, then $w(t)$ is an
eigenvalue of $A\left(\kappa;\mu(t),\nu(t)\right)$ for each $t\in\mathfrak{D}$. Furthermore, if $\kappa$ is a
half-integer and $w(t_0)$ is a monodromy eigenvalue of $A\left(\kappa;\mu(t_0),\nu(t_0)\right)$, then $w(t)$ is a
monodromy eigenvalue of $A\left(\kappa;\mu(t),\nu(t)\right)$ for each $t\in\mathfrak{D}$.
\end{lemma}

\begin{proof}
In terms of \eqref{Coords} and $\lambda(t) := w(t)$, the coefficient matrices \eqref{Coeff1} of the system \eqref{AngSys}
take the form
\begin{gather*}
B_0(t) = \left(\begin{array}{cc} -\frac{\kappa}{2}-\frac{1}{4} & \frac{t}{2}\left(v(t)+\frac{\sigma}{v(t)}\right) - w(t) \\[1ex] 0 &  \frac{\kappa}{2}+\frac{1}{4} \end{array}\right),\quad
B_1(t) = \left(\begin{array}{cc}  \frac{\kappa}{2}+\frac{1}{4} & 0 \\[1ex] \frac{t}{2}\left(v(t)+\frac{\sigma}{v(t)}\right) - w(t) & -\frac{\kappa}{2}-\frac{1}{4} \end{array}\right),\\
 C(t)  = \frac{t}{v(t)}\left(\begin{array}{rr} -\left(v(t)^2+1\right) & -\left(v(t)^2-1\right) \\[2ex]
                                                \left(v(t)^2-1\right) &  \left(v(t)^2+1\right) \end{array}\right),
\end{gather*}
where the condition \Hyp{i} is satisfied with $\alpha=-\frac{\kappa}{2}-\frac{1}{4}$ and
$\beta=\frac{\kappa}{2}+\frac{1}{4}$. Now, if we define the matrix function
\begin{equation*}
\Omega(x,t) := \frac{1}{v(t)}\left(\begin{array}{cc}
\left(v(t)^2-\sigma\right)\left(\frac{1}{2}-x\right) & \left(v(t)^2+\sigma\right)(1-x) \\[2ex]
\left(v(t)^2+\sigma\right)x & \left(v(t)^2-\sigma\right)\left(x-\frac{1}{2}\right)
\end{array}\right),\quad (x,t)\in\C\times\mathfrak{D},
\end{equation*}
then, by a straightforward calculation using the characteristic equations \eqref{Char1} -- \eqref{Char2}, it follows that
the deformation equation in \Hyp{iii} holds. Finally, by setting
\begin{equation*}
G_0(t) := \left(\begin{array}{cc} e^{\phi(t)} & \left[\frac{t}{2}\left(v(t)+\frac{\sigma}{v(t)}\right)-w(t)\right]e^{-\phi(t)} \\[2ex] 0 & \left(\kappa+\frac{1}{2}\right)e^{-\phi(t)} \end{array}\right),\quad
G_1(t) := K G_0(t) K,
\end{equation*}
where $K$ is the matrix \eqref{K} and
\begin{equation*}
\phi(t) := \int_{t_0}^t \frac{v(\tau)^2-\sigma}{2\,v(\tau)}\,\d\tau,\quad t\in\mathfrak{D},
\end{equation*}
the conditions \Hyp{ii} and \Hyp{iv} are satisfied. Since a point $w(t)$ is a monodromy eigenvalue of
$A\left(\kappa;\mu(t),\nu(t)\right)$ if and only if \eqref{AngSys} has the property \Hyp{P}, and $w(t)$ is a classical
eigenvalue of $A\left(\kappa;\mu(t),\nu(t)\right)$ if and only if \eqref{AngSys} has the property \Hyp{H}, the assertion
follows from Corollary \ref{Isomono}.
\end{proof}

\begin{theorem} \label{MEV2}
For a fixed $\kappa=k-\frac{1}{2}$ with a positive integer $k$, let $(0,0)\in\mathfrak{S}\subset\R^2$ be a simply
connected domain such that for each $(\mu,\nu)\in\mathfrak{S}$ all monodromy eigenvalues $\lambda_0^j(\kappa;\mu,\nu)$,
$j=1-k,\ldots,k-1$, of $A(\kappa;\mu,\nu)$ are simple zeros of the polynomial $P(\kappa;\sdot,\mu,\nu)$ given by Theorem
\ref{MEV1}. Then each function $\lambda=\lambda_0^j$, $j=1-k,\ldots,k-1$, satisfies the partial differential equation
\eqref{PDE} in $\mathfrak{S}$.
\end{theorem}

\begin{proof}
Let $j\in\{1-k,\ldots,k-1\}$ be fixed. The monodromy eigenvalues of $A(\kappa;\mu,\nu)$ are exactly the zeros of
the polynomial $P(\kappa;\sdot;\mu,\nu)$, and since all zeros of $P(\kappa;\sdot;\mu,\nu)$ are simple, the implicit
function theorem implies that $\lambda_0^j(\kappa;\mu,\nu)$ depends analytically on $(\mu,\nu)$ in $\mathfrak{S}$. In order
to show that the function $\lambda=\lambda_0^j$ satisfies the PDE \eqref{PDE}, we make use of the unique continuation
property of analytical functions. That means, it suffices to prove that \eqref{PDE} holds for $\lambda=\lambda_0^j$ in a
neighbourhood of some point $(\mu,\nu)=(\tau,0)\in\mathfrak{S}$, $\tau>0$. Now, in view of the coordinate transformation
\eqref{Coords}, we have to verify that the function $\lambda_0^j\left(\kappa;\mu(t,v),\nu(t,v)\right)$ is a solution of
the partial differential equation \eqref{PDEw} in a neighbourhood of the point $(t,v)=(\tau,1)$. To this end, let us
consider the characteristic equations of \eqref{PDEw}
\begin{equation*}
\frac{\partial v}{\partial t}(t,u) = -\frac{2\,v(t,u)\,w(t,u)}{t},\quad
\frac{\partial w}{\partial t}(t,u) = -\kappa\left(v(t,u)+\frac{1}{v(t,u)}\right) - \frac{t}{2}\left(v(t,u)^2-\frac{1}{v(t,u)^2}\right)
\end{equation*}
together with the initial values
\begin{equation*}
v(\tau,u) = u,\quad w(\tau,u) = \lambda_0^j\left(\kappa;\mu(\tau,u),\nu(\tau,u)\right),
\end{equation*}
which depend analytically on the parameter $u\in(0,\infty)$. The solutions $v(t,u)$ and $w(t,u)$ of this
initial value problem are analytical functions in a neighbourhood of $(\tau,1)$, and since
$\frac{\partial v}{\partial u}(\tau,u)=1$, they form locally an integral surface for the PDE \eqref{PDEw}
(compare \cite[Chap. 1, Sec. 5]{John}). More precisely, there exists an analytical function $U$ defined on a neighbourhood
$\mathfrak{V}$ of $(t,v)=(\tau,1)$ such that $U(\tau,v)=v$, and $W(t,v) := w\left(t,U(t,v)\right)$ is a solution of
\eqref{PDEw} in $\mathfrak{V}$. Now, Lemma \ref{MPD} implies that $W(t,v)$ is a monodromy eigenvalue of
$A\left(\kappa;\mu(t,v),\nu(t,v)\right)$ for all $(t,v)\in\mathfrak{V}$, and since
$W(\tau,v)=\lambda_0^j\left(\kappa;\mu(\tau,v),\nu(\tau,v)\right)$, it follows that
$W(t,v)=\lambda_0^j\left(\kappa;\mu(t,v),\nu(t,v)\right)$ holds identically on $\mathfrak{V}$. This completes the proof of
the Theorem.
\end{proof}

In a similar way we can apply Lemma \ref{MPD} to prove that for fixed $\kappa\in(0,\infty)$ the zeros of the
function $\lambda\longmapsto\Delta(\kappa;\lambda,\mu,\nu)$ defined in Section II and therefore the eigenvalues of
$A(\kappa;\mu,\nu)$ satisfy the partial differential equation \eqref{PDE}. This alternative proof of Theorem \ref{KPT} is
based on monodromy preserving deformation -- a general technique, which should be applicable to other eigenvalue problems
as well. Potential candidates and associated $\Omega$-matrices for solving the deformation equations can be found in
\cite[Appendix C]{JMU2}.

Finally, as a consequence of Theorem \ref{MEV2}, the zeros of the polynomial $P(\kappa;\sdot;\mu,\nu)$ given by
Theorem \ref{MEV1} satisfy the PDE \eqref{PDE} and do not coincide with any eigenvalue of $A(\kappa;\mu,\nu)$ in a
neighbourhood of $(\mu,\nu)=(0,0)$. Moreover (see the proof of Lemma \ref{AlgSol}), $P(\kappa;\sdot;\mu,\nu)$ gives rise
to a special integral of polynomial type for the Painlev{\'e} III \eqref{PIII}. Now, the results of Mansfield \& Webster
in \cite[Section 2]{MW} suggest that these special integrals are unique in some sense, which in turn implies that
classical eigenvalues of the Chandrasekhar-Page angular equation are not algebraic.

\section*{Appendix}

\subsection{Eigenvalues and eigenfunctions in the case $\mu=\nu=0$}

For fixed $\kappa\in[\frac{1}{2},\infty)$, a point $\lambda\in\C$ is an eigenvalue of $A(\kappa;0,0)$ if and only if the system
\eqref{AngEqu} with $(\mu,\nu)=(0,0)$ has a nontrivial solution $S(\theta)$ satisfying
\begin{equation} \label{AngInt0}
\int_{0}^{\pi} |S(\theta)|^2\,\d\theta < \infty.
\end{equation}
Introducing the functions $u,\,v:(-1,1)\longrightarrow\C$ by
\begin{equation} \label{Trafo0}
S(\theta) =:
\sin^{\kappa+\frac{1}{2}}\theta\left(\begin{array}{c}
\sqrt{\tan\frac{\theta}{2}}\,u(\cos\theta) \\[1ex]
\sqrt{\cot\frac{\theta}{2}}\,v(\cos\theta) \end{array}\right),
\end{equation}
then \eqref{AngEqu} with $(\mu,\nu)=(0,0)$ is transformed into
\begin{equation} \label{Diff0}
(1-x)u'(x) = \left(\kappa+\frac{1}{2}\right)u(x) + \lambda\,v(x),\quad
(1+x)v'(x) = -\lambda\,u(x) - \left(\kappa+\frac{1}{2}\right)v(x),
\end{equation}
and the normalisation condition \eqref{AngInt0} is equivalent to
\begin{equation} \label{Norm0}
\int_{-1}^{1} u(x)^2(1-x)^{\kappa+\frac{1}{2}}(1+x)^{\kappa-\frac{1}{2}}\,\d x < \infty,\quad
\int_{-1}^{1} v(x)^2(1-x)^{\kappa-\frac{1}{2}}(1+x)^{\kappa+\frac{1}{2}}\,\d x < \infty.
\end{equation}
If $\lambda = 0$, then the differential equations \eqref{Diff0} imply
$u(x)=c_1(1-x)^{-\kappa-\frac{1}{2}}$ and $v(x)=c_2(1+x)^{-\kappa-\frac{1}{2}}$ with some constants $c_1,\,c_2\in\C$,
and from the condition \eqref{Norm0} it follows that $c_1=c_2=0$. Hence, $\lambda=0$ is not an eigenvalue of
$A(\kappa;0,0)$, and we assume in what follows that $\lambda\neq 0$. In this case, the second equation in \eqref{Diff0}
gives
\begin{equation} \label{Elim}
u(x) = -\frac{1+x}{\lambda}\,v'(x)-\frac{\kappa+\frac{1}{2}}{\lambda}\,v(x),
\end{equation}
and for $v$ we obtain the second order differential equation
\begin{equation*}
(1-x^2)v''(x) + \left[1-2(\kappa+1)x\right]v'(x)+\left[\lambda^2-\left(\kappa+\frac{1}{2}\right)^2\right]v(x)=0.
\end{equation*}
If we set $\alpha := \kappa-\frac{1}{2}$, $\beta := \kappa+\frac{1}{2}$ and $\Lambda := \lambda-\kappa-\frac{1}{2}$,
this differential equation becomes
\begin{equation} \label{Jacobi}
(1-x^2)v''(x) + \left[\beta-\alpha-(\alpha+\beta+2)x\right]v'(x)+\Lambda(\Lambda+\alpha+\beta+1)v(x) = 0,
\end{equation}
and the second condition in \eqref{Norm0} takes the form
\begin{equation} \label{Weight}
\int_{-1}^{1} v(x)^2(1-x)^\alpha(1+x)^\beta\,\d x < \infty.
\end{equation}
Note that \eqref{Jacobi} and \eqref{Weight} is the eigenvalue problem associated to the Jacobi polynomials. More
precisely, the solutions of the differential equation \eqref{Jacobi} which are square integrable with respect to the
weight function $(1-x)^\alpha(1+x)^\beta$ are constant multiples of the Jacobi Polynomials $P_n^{(\alpha,\beta)}$ with
some non-negative integer $n$, and the corresponding eigenvalues $\lambda_n^{\pm}$ are determined by the equation
$\lambda^2-\left(\kappa+\frac{1}{2}\right)^2=n(n+\alpha+\beta+1)$, i.e.,
$\lambda_n^{\pm} = \pm\left(\kappa+\frac{1}{2}+n\right)$. Now, if we define $v(x) := -P_n^{(\alpha,\beta)}(x)$,
$x\in(-1,1)$, then \eqref{Elim} yields
\begin{align*}
\lambda_n^\pm\, u(x)
& = (1+x)\frac{\d}{\d x}P_n^{(\alpha,\beta)} + \beta\,P_n^{(\alpha,\beta)}
  = \frac{\alpha+\beta+n+1}{2}(1+x)P_{n-1}^{(\alpha+1,\beta+1)} + \beta\,P_n^{(\alpha,\beta)} \\
& = \frac{\alpha+\beta+n+1}{\alpha+\beta+2n+1}\left[(\beta+n)P_{n-1}^{(\alpha+1,\beta)} + n\,P_n^{(\alpha+1,\beta)}\right] + \beta\,P_n^{(\alpha,\beta)} \\
& = (\alpha+\beta+n+1)P_n^{(\alpha+1,\beta)} - (\alpha+n+1)P_n^{(\alpha,\beta)} = (\beta+n)P_n^{(\alpha+1,\beta-1)} = |\lambda_n^{\pm}|P_n^{(\alpha+1,\beta-1)}
\end{align*}
where we applied the differentiation formulas and contiguous relations for Jacobi polynomials
(see \cite[Section 5.2]{MOS}). Hence, $u(x) = \pm P_n^{(\alpha+1,\beta-1)}(x)$, $x\in(-1,1)$, and since $u$ satisfies the
first condition in \eqref{Norm0}, the numbers $\lambda_n^{\pm}$ are in fact eigenvalues of $A(\kappa;0,0)$. Moreover,
the corresponding eigenfunctions are constant multiples of
\begin{equation*}
\sin^\kappa\theta\left(\begin{array}{r}
\pm\sqrt{\tan\frac{\theta}{2}}\,P_n^{(\kappa+\frac{1}{2},\kappa-\frac{1}{2})}(\cos\theta) \\[1ex]
 - \sqrt{\cot\frac{\theta}{2}}\,P_n^{(\kappa-\frac{1}{2},\kappa+\frac{1}{2})}(\cos\theta)
\end{array}\right),\quad\theta\in(0,\pi),
\end{equation*}
which form a complete orthogonal set in $\hs^2\left((0,\pi),\C^2\right)$. In particular, the spectrum of $A(\kappa;0,0)$
is given by $\{\lambda_n^{\pm}:n=0,1,2,\ldots\}$.

\subsection{A Numerical Example}

As a numerical example, we have computed the coefficients $c_{m,n}$ of the power series expansion \eqref{Series} up to
and including $m+n=8$ for $\kappa=\frac{1}{2}$ and $j=1$ using the recurrence relation given in Section III. The
coefficients have been rounded to six significant figures and listed in the table below. It should be noted that they are
to some extent different from the coefficients displayed in \cite[Table I]{SFC}. Evaluating the power series expansion
\eqref{Series} at $\alpha=0.01$ and $\beta=0.02$, i.e., $(\mu,\nu)=(0.005,0.015)$, yields
$\tilde\lambda_1 = \mathtt{1.01167}$ as a numerical approximation for the eigenvalue $\lambda_1$, and this result
coincides with the value given in \cite[Table II]{SFC}. For a second pair of parameters $(\alpha,\beta)=(0.5,1.0)$, i.e.,
$(\mu,\nu)=(0.25,0.75)$, we obtain $\tilde\lambda_1 = \mathtt{1.59745}$, which differs slightly from the value
$\hat\lambda_1=\mathtt{1.59764}$ listed in \cite[Table II]{SFC}. In order to test the reliability of our numerical
result, we can use the statement of Lemma \ref{Zeros2}. That means, we approximate $\Theta(\lambda)$ defined in
\eqref{Theta} by the second component $\Theta_n(\lambda)$ of $d_n(\lambda)$ for $n=8$, and we compare
$\Theta_8(\tilde\lambda_1)$ and $\Theta_8(\hat\lambda_1)$ with the theoretical result $\Theta(\lambda_1)=0$. As
$\Theta_8(\tilde\lambda_1)=\mathtt{3.60882e-05}$ and $\Theta_8(\hat\lambda_1)=\mathtt{-2.51164e-04}$, our result seems
to be more trustworthy. Finally, let $(\mu,\nu)=(0.02,0.1)$. The coefficients of the polynomial $\Theta_8$ are given in
Table \ref{Table2}. For these parameters, our power series approximation gives $\tilde\lambda_1=\mathtt{1.07379}$ which
differs significantly from the value $\hat\lambda_1=\mathtt{1.06104}$ given by Chakrabarti (see
\cite[Table 1]{Chakrabarti}). Despite his claiming of an accuracy of six decimals, the evaluation
of $\Theta_8$ at the eigenvalues in question gives $\Theta_8(\tilde\lambda_1)=\mathtt{5.68899e-12}$ and
$\Theta_8(\hat\lambda_1)=\mathtt{1.52770e-02}$ in favour of our result. Thus, Chakrabarti's calculations should
be taken with some caution.

\begin{table}[h]
\ttfamily\small\setlength{\tabcolsep}{5pt}
\begin{tabular}{r|rrrrrrrrr}
  & $m =$ \hfill $0$ & $1$ & $2$ & $3$ & $4$ & $5$ & $6$ & $7$ & $8$ \\ \hline
$n = 0$ &  1.00000e+00 &  5.00000e-01 &  0.00000 &  0.00000 &  0.00000 &  0.00000 &  0.00000 &  0.00000 &  0.00000 \\
$1$ &  1.66667e-01 &  0.00000 &  0.00000 &  0.00000 &  0.00000 &  0.00000 &  0.00000 &  0.00000 &  \\
$2$ &  7.40741e-02 & -1.48148e-02 &  0.00000 &  0.00000 &  0.00000 &  0.00000 &  0.00000 &  &  \\
$3$ & -8.23045e-03 &  3.29218e-03 & -4.70312e-04 &  0.00000 &  0.00000 &  0.00000 &  &  &  \\
$4$ & -9.14495e-04 &  5.48697e-04 & -1.22281e-04 &  1.35868e-05 &  0.00000 &  &  &  &  \\
$5$ &  5.08053e-04 & -4.06442e-04 &  1.41790e-04 & -2.67091e-05 &  &  &  &  &  \\
$6$ & -3.38702e-05 &  3.38702e-05 & -1.63351e-05 &  &  &  &  &  &  \\
$7$ & -2.63435e-05 &  3.16122e-05 &  &  &  &  &  &  &  \\
$8$ &  7.10856e-06 &  &  &  &  &  &  &  &  \\
\end{tabular}
\caption{The coefficients $c_{m,n}$, $0\leq m+n\leq 8$, of the power series expansion \eqref{Series}
in the case $\kappa=\frac{1}{2}$ and $j=1$.}
\end{table}

\begin{table}
\ttfamily\small\setlength{\tabcolsep}{5pt}
\begin{tabular}{r|r||r|r}
$n= 0$ &  1.22151e+00 & $n = 9$ &  4.91151e-06 \\
$ 1$ &  1.44347e-02 & $ 10$ & -4.22048e-04 \\
$ 2$ & -1.70525e+00 & $ 11$ & -9.46610e-08 \\
$ 3$ & -7.92297e-03 & $ 12$ &  1.02643e-05 \\
$ 4$ &  6.72114e-01 & $ 13$ &  6.88933e-10 \\
$ 5$ &  1.46003e-03 & $ 14$ & -1.26470e-07 \\
$ 6$ & -1.12351e-01 & $ 15$ & -1.00000e-26 \\
$ 7$ & -1.21028e-04 & $ 16$ &  6.15119e-10 \\
$ 8$ &  9.39664e-03 & \multicolumn{2}{c}{}
\end{tabular}
\caption{The coefficients $\delta_n$ of the polynomial $\Theta_8(\lambda)=\sum_{n=0}^{16}\delta_n \lambda^n$
for $\kappa=\frac{1}{2}$, $\mu=0.02$, $\nu=0.1$. \label{Table2}}
\end{table}

\subsection{Eigenfunctions in the case $|\mu|\neq |\nu|$}

Eliminating the second component of $y$ in the system \eqref{AngSys}, we get a linear second-order differential equation
for the first component $y_1$ given by
\begin{equation} \label{preHeun}
\frac{d^{2}y_1}{dx^{2}}(x)+\left(\frac{1}{x}-\frac{1}{x-b}\right)\frac{dy_1}{dx}(x)+\left(\tau_{0}+\frac{\tau_{1}}{x}+\frac{\tau_2}{x^2}+\frac{\tau_3}{x-1}+\frac{\tau_{4}}{(x-1)^{2}}+\frac{\tau_5}{x-b}\right)y_1(x)=0
\end{equation}
with
\begin{equation*}
b:=\frac{\mu-\lambda}{2\,\mu},\quad \tau_{0}:=4\left(\mu^{2}-\nu^{2}\right), \quad
\tau_{1}:=\lambda^{2}-2\,\alpha^{2}+2\,\nu+\alpha-\mu^2-4\,\alpha\,\nu+\frac{2\,\alpha\,\mu}{\mu-\lambda}, \quad
\alpha:=\frac{\kappa}{2}+\frac{1}{4}
\end{equation*}
and
\begin{equation*}
\tau_{2}:=-\alpha^2, \quad
\tau_{3}:=\frac{4\,\alpha\,\mu^2}{\mu^2-\lambda^2}+2\,\nu-\tau_1, \quad \tau_{4}:=\alpha(1-\alpha), \quad
\tau_{5}:=\frac{2(\nu\,\mu^2+2\,\alpha\,\mu^2-\nu\,\lambda^2)}{\lambda^2-\mu^2}.
\end{equation*}
Now by means of the transformation
\[
y_1(x) := x^\alpha(x-1)^\alpha\psi(x)e^{2tx}, \quad t=\pm\sqrt{\nu^{2}-\mu^{2}},
\]
we find that $\psi(x)$ satisfies the generalised Heun equation
\begin{equation} \label{Heun}
\frac{d^{2}\psi(x)}{dx^{2}}+\left(\frac{1-\mu_{0}}{x}+\frac{1-\mu_{1}}{x-1}+\frac{1-\mu_{2}}{x-b}+4\,t\right)\frac{d\psi(x)}{dx}+\frac{\beta_{0}+\beta_{1}x+\beta_{2}x^{2}}{x(x-1)(x-b)}\psi(x)=0,
\end{equation}
where
\begin{equation*}
\mu_{0}=-2\,\alpha, \quad \mu_{1}=1-2\,\alpha, \quad \mu_{2}=2, \quad \beta_{2}:=8\,\alpha\,t,
\end{equation*}
and
\begin{align*}
\beta_{1} & = \mu^2-\lambda^2 -2\,t\left[b+2\,\alpha(1+2\,b)\right]+2\,\alpha(\alpha-1)+2\,\nu(2\,\alpha-b)
              - \frac{2\,\alpha\,\mu(b-1)}{\lambda+\mu} + \frac{2\,\alpha\,\mu\,b}{\mu-\lambda}, \\
\beta_{0} & = b\,(\lambda^2-\mu^2) + b\left[2(\nu+t)-4\,\alpha(\nu-t)-4\,\alpha^2\right]+\alpha-\frac{2\,\mu\,\alpha\,b}{\lambda-\mu}.
\end{align*}
We observe that $0$, $1$ and $b$ are simple singularities with characteristic exponents $(0,\mu_{0})$, $(0,\mu_{1})$ and
$(0,\mu_{2})$ respectively, while $\infty$ is (at most) an irregular singularity of rank $1$. To stress the importance of
equation \eqref{Heun}, it is sufficient to remark that it contains the ellipsoidal wave equation as well as Heun's
equation and thus the Mathieu, spheroidal, Lam\'{e}, Whittaker-Hill and Ince equations as special cases.

\section*{Acknowledgement}

\noindent
M.~Winklmeier gratefully acknowledges the support of the German Research Foundation, DFG, Grant {No.~TR\,368/4--1}, and
D.~Batic is indebted to the financial support of the MPI f{\"u}r Math. i. d. Naturw., Leipzig, Germany. The authors also
thank Felix Finster, Universit{\"a}t Regensburg, Germany, and Christiane Tretter, Universit{\"a}t Bremen, Germany, for
fruitful discussions. Finally, they thank Alexander Kitaev, Steklov Mathematical Institute, St. Petersburg, Russia, for
suggestions on literature about Painlev{\'e} III.

\end{document}